\documentclass[10pt,twocolumn,twoside]{IEEEtran}

\usepackage{amssymb}
\usepackage{dsfont}
\usepackage{mathdots}
\usepackage{amsmath}
\usepackage{graphicx}
\usepackage{dcolumn}
\usepackage{bm}
\usepackage{bbm}
\usepackage{color}
\usepackage{dcolumn}
\usepackage{tcolorbox}
\usepackage{color}
\usepackage{theorem}
\usepackage{times,amsmath,epsfig}
\usepackage{amssymb}
\usepackage{subfigure}
\usepackage{cite}
\usepackage{cases}
\usepackage{mathtools}
\usepackage{multirow}
\usepackage{booktabs}

\usepackage{algorithmic}
\usepackage{algorithm}
\usepackage{needspace}

\usepackage{tikz}
\usetikzlibrary{shapes,arrows}

\newenvironment{myproof}[1][$\!\!$]{{\noindent\bf Proof #1: }}
                         {\hfill$\blacksquare$\medskip}

\newcounter{excercise}
\newcounter{excercisepart}

\definecolor{pennblue}{cmyk}{1,0.65,0,0.30}
\definecolor{pennred}{cmyk}{0,1,0.65,0.34}
\definecolor{mygreen}{rgb}{0.10,0.50,0.10}

\newcommand \blue[1]        {{\color{black}#1}}

\newcommand   \ind   [1] {{\mathbb I \left\{#1\right\}  } }

\def \reals    {{\mathbb R}}

\newcommand{\argmax}{\operatornamewithlimits{argmax}}

\def\ccalA{{\ensuremath{\mathcal A}}}
\def\ccalB{{\ensuremath{\mathcal B}}}

\def\ccalF{{\ensuremath{\mathcal F}}}
\def\ccalG{{\ensuremath{\mathcal G}}}

\def\ccalI{{\ensuremath{\mathcal I}}}

\def\ccalM{{\ensuremath{\mathcal M}}}

\def\ccalV{{\ensuremath{\mathcal V}}}
\def\ccalS{{\ensuremath{\mathcal S}}}
\def\ccalT{{\ensuremath{\mathcal T}}}

\def\ccal0{{\ensuremath{\mathcal 0}}}
\def\tdx{{\ensuremath{\tilde x}}}

\def\bbA{{\ensuremath{\mathbf A}}}
\def\bbB{{\ensuremath{\mathbf B}}}

\def\bbD{{\ensuremath{\mathbf D}}}

\def\bbG{{\ensuremath{\mathbf G}}}
\def\bbH{{\ensuremath{\mathbf H}}}
\def\bbI{{\ensuremath{\mathbf I}}}
\def\bbJ{{\ensuremath{\mathbf J}}}

\def\bbL{{\ensuremath{\mathbf L}}}
\def\bbM{{\ensuremath{\mathbf M}}}

\def\bbP{{\ensuremath{\mathbf P}}}

\def\bbR{{\ensuremath{\mathbf R}}}

\def\bbU{{\ensuremath{\mathbf U}}}
\def\bbV{{\ensuremath{\mathbf V}}}

\def\bbX{{\ensuremath{\mathbf X}}}

\def\bbg{{\ensuremath{\mathbf g}}}
\def\bbh{{\ensuremath{\mathbf h}}}

\def\bbn{{\ensuremath{\mathbf n}}}

\def\bbu{{\ensuremath{\mathbf u}}}
\def\bbv{{\ensuremath{\mathbf v}}}

\def\bbx{{\ensuremath{\mathbf x}}}
\def\bby{{\ensuremath{\mathbf y}}}

\def\bb0{{\ensuremath{\mathbf 0}}}
\def\hbU{{\hat{\ensuremath{\mathbf U}} }}

\def\tbx{{\tilde{\ensuremath{\mathbf x}} }}
\def\tby{{\tilde{\ensuremath{\mathbf y}} }}

\def\bbpi{\boldsymbol{\pi}}

\newcommand{\RN}[1]{%
	\textup{\uppercase\expandafter{\romannumeral#1}}%
}
\newcommand{\rn}[1]{%
	\textup{\lowercase\expandafter{\romannumeral#1}}%
}

\newcommand{\abs}[1]{\lvert #1 \rvert}

\newtheorem{mytheorem}{\bf Theorem}
\newtheorem{mydefinition}{\bf Definition}

\newtheorem{mylemma}{\bf Lemma}
\newtheorem{myproposition}{\bf Proposition}
\newtheorem{remark}{\bf Remark}

\def\lovasz{Lov\'asz }

\newtcolorbox{myblockt}[1]{colback=urblue!5!white,
	colframe=urblue,fonttitle=\bfseries,
	title=#1}

\newtcolorbox{myblock}{colback=urblue!5!white,
	colframe=urblue,fonttitle=\bfseries}

\title{A Directed Graph Fourier Transform\\ with Spread Frequency Components}
\author{\IEEEauthorblockN{Rasoul Shafipour, \emph{Student Member, IEEE}, Ali Khodabakhsh, \emph{Student Member, IEEE},\\ Gonzalo Mateos, \emph{Senior Member, IEEE}, and Evdokia Nikolova}
\thanks{Work in this paper was supported by the NSF awards CCF-1750428, CCF-1216103, CCF-1331863, CCF-1350823, and CCF-1733832. R. Shafipour and G. Mateos are with the Dept. of Electrical and Computer Engineering, University of Rochester. A. Khodabakhsh and E. Nikolova are with the Dept. of Electrical and Computer Engineering, University of Texas at Austin. Emails: rshafipo@ece.rochester.edu, ali.kh@utexas.edu, gmateosb@ece.rochester.edu, and nikolova@austin.utexas.edu. Part of the results in this paper were presented at the Fifth IEEE Global Conference on Signal and Information Processing, Montreal, Canada, November 2017~\cite{shafipour2017digraph} and at the Fourty-Third IEEE International Conference on Acoustics, Speech, and Signal Processing, Calgary, Canada, April 2018~\cite{shafipour2018digraph}.}}

\begin{document}
\maketitle

\begin{abstract}%
We study the problem of constructing a graph Fourier transform (GFT) for directed graphs (digraphs), which decomposes graph signals into different modes of variation with respect to the underlying network. Accordingly, to capture low, medium and high frequencies we seek a digraph (D)GFT such that the orthonormal frequency components are as spread as possible in the graph spectral domain. To that end, we advocate a two-step design whereby we: (\romannumeral 1)  find the maximum directed variation (i.e., a novel notion of frequency on a digraph) a candidate basis vector can attain; and (\romannumeral 2) minimize a smooth spectral dispersion function over the achievable frequency
range to obtain the desired spread DGFT basis. Both steps involve non-convex, orthonormality-constrained optimization problems, 
 {which are efficiently tackled via a feasible optimization method on the Stiefel manifold that provably converges to a stationary solution.}
We also propose a heuristic to construct the DGFT basis from Laplacian eigenvectors of an undirected version of the digraph. We show that the spectral-dispersion minimization problem can be cast as supermodular optimization over the set of candidate frequency components, whose orthonormality can be enforced via a matroid basis constraint. This motivates adopting a scalable greedy algorithm to obtain an approximate solution with quantifiable worst-case spectral dispersion. We illustrate the effectiveness of our DGFT algorithms through numerical tests on synthetic and real-world networks. We also carry out a graph-signal denoising task, whereby the DGFT basis is used to decompose and then low-pass filter temperatures recorded across the United States.

\vspace{0.2cm}
\end{abstract}


%

\section{Introduction}\label{S:Introduction}
Network processes such as neural activities at different regions of the brain \cite{honey2007network}, vehicle trajectories over road networks~\cite{deri_nyc_taxi}, or, infectious states of individuals in a population affected by an epidemic \cite{kolaczyk2009statistical}, can be represented as graph signals supported on the vertices of the adopted graph abstraction to the network. Under the natural assumption that the signal properties relate to the underlying graph (e.g., when observing a network diffusion or percolation process), the goal of graph signal processing (GSP) is to develop algorithms that fruitfully exploit this relational structure \cite{gsp2018tutorial,sandryhaila2013}. {From this vantage point,} generalizations of traditional signal processing tasks such as filtering \cite{sandryhaila2013,teke2017extending,tremblay2017design,isufi2017autoregressive,segarra2017filters}, sampling and reconstruction \cite{marques2016sampling,chen2015discrete}, statistical GSP and spectrum estimation \cite{marques2016stationary,girault2015stationary,perraudin2017stationary}, {(blind)} filter identification \cite{rasoul,segarra2017blindid} as well as signal representations \cite{thanou2014learning,zhu2012approximating,shuman_graph_wavelts} have been recently explored under the GSP purview~\cite{gsp2018tutorial}.

An instrumental GSP tool is the graph Fourier transform (GFT), which decomposes a graph signal into orthonormal components describing different modes of variation with respect to the graph topology. The GFT provides a method to equivalently represent a graph signal in two different domains -- the graph domain, consisting of the nodes in the graph, and the frequency domain, represented by the frequency  \blue{components} of the graph. Therefore, signals can be manipulated in the frequency domain to induce different levels of interactions between neighbors in the network. Here we aim to generalize the GFT to directed graphs (digraphs); see also \cite{DSP_freq_analysis,sardellitti}. We first propose a novel notion of signal variation (frequency) over digraphs and find an approximation of the maximum possible frequency ($f_{\max}$) that a unit-norm graph signal can achieve. We design a digraph {(D)GFT} such that the resulting frequencies (i.e., the directed variation of the sought orthonormal  \blue{basis vectors}) distribute as evenly as possible across $[0,f_{\max}]$. Beyond offering parsimonious representations of slowly-varying signals on digraphs, a DGFT with spread frequency components can facilitate more interpretable frequency analyses and aid filter design in the spectral domain. In a way, to achieve these goals we advocate a form of regularity in the DGFT-induced frequency domain. A different perspective is to consider an irregular (dual) graph support, which as argued in~\cite{leus2017dualshift} can offer complementary merits and insights.

\noindent \textbf{Related work.} To position our contributions in context, we first introduce some basic GSP notions and terminology. Consider a weighted digraph $\ccalG=(\ccalV,\bbA)$, where $\ccalV$ is the set of nodes {(i.e., vertices)} with cardinality $\lvert \ccalV \rvert=N$, and $\bbA \in \reals^{N \times N}$ is the graph adjacency matrix with entry $A_{ij}$ denoting the edge weight from node $i$ to  node $j$.
We assume that the edge weights in $\ccalG$ are non-negative ($A_{ij} \geq 0$). 
For an undirected graph\blue{,} $\bbA$ is symmetric, and the positive semi-definite Laplacian matrix takes the form $\bbL \coloneqq \bbD - \bbA$, where $\bbD$ is the diagonal degree matrix with $D_{ii} = \sum_{j} A_{ji}$. A graph signal $\bbx : \ccalV \mapsto \reals^N$ can be represented as a vector of length $N$, where entry $x_i$ {denotes} the signal value at node $i \in \ccalV$.

For undirected graphs, the GFT of signal $\bbx$ is often defined as $\tilde{\bbx} = \bbV^T \bbx$, where $\bbV \coloneqq [\bbv_1, \ldots , \bbv_N]$ comprises the eigenvectors of the Laplacian~\cite{gsp2018tutorial,tremblay2017design,zhu2012approximating,girault2018gft}. Interestingly, in this setting the GFT encodes a notion of signal variability over the graph akin to the notion of frequency in Fourier analysis of temporal signals. To understand this analogy, define the total variation of the signal $\bbx$ with respect to the Laplacian $\bbL$ as
\begin{equation} \label{e:TV_def}
\text{TV}(\bbx) = \bbx^T \bbL \bbx = \sum_{i,j=1,j>i}^{N} A_{ij} (x_i - x_j)^2.
\end{equation}
The total variation $\text{TV}(\bbx)$ is a smoothness measure, quantifying how much the signal $\bbx$ changes with respect to the expectation on variability that is encoded by the weights $\bbA$. Consider the total variation of the eigenvectors $\bbv_k$, which is given by $\text{TV}(\bbv_k) = \lambda_k$, the $k$th Laplacian eigenvalue. Hence, eigenvalues {$0=\lambda_1<\lambda_2\leq\ldots\leq \lambda_N$} can be viewed as graph frequencies, indicating how the GFT \blue{basis vectors} (i.e., the frequency components) vary over the graph $\ccalG$. Note that there may be more than one eigenvector corresponding to a graph frequency in case of having repeated eigenvalues.  Moreover, frequency components associated with close eigenvalues can often be quite dissimilar and focus on different parts of the graph~\cite{teke2017uncertainty}.

Extensions of the combinatorial Laplacian to digraphs have also been proposed~\cite{chung2005laplacians,singh2016graph}. 
However, eigenvectors of the directed Laplacian  generally fail to yield spread frequency components as we illustrate in Section \ref{S:numerical}.
A more general GFT definition is based on the Jordan decomposition of adjacency matrix $\bbA=\bbV \bbJ \bbV^{-1}${,} where the frequency representation of graph signal $\bbx$ is $\tilde{\bbx} = \bbV^{-1} \bbx$ \cite{DSP_freq_analysis}. While valid for digraphs, the associated notion of signal variation in \cite{DSP_freq_analysis} does not ensure that constant signals have zero variation. Moreover, $\bbV$ is not necessarily orthonormal and Parseval's identity does not hold. {From a computational standpoint}, obtaining the Jordan decomposition is {often} numerically unstable; see also \cite{deri2017spectral}  \blue{and~\cite{girault2015signal} for recent stabilizing alternatives.}  
 \blue{On a related note, a class of energy-preserving graph-shift operators with desireable properties are constructed in~\cite{gavili2017ongso}. Starting from the adjacency matrix of an arbitrary graph, the idea is to preserve eigenvectors (hence resulting in the same GFT as in~\cite{DSP_freq_analysis}) and replace the eigenvalues with pure phase shifts.}
Recently, a fresh look at the GFT for digraphs was put forth in \cite{sardellitti} based on minimization of {the} (convex) \lovasz extension of the graph cut size, subject to orthonormality constraints on the desired  \blue{basis}. 
While the GFT \blue{basis vectors} in \cite{sardellitti} tend to be constant across clusters of the graph, in general they may fail to yield signal representations capturing different modes of signal variation with respect to $\ccalG$; see Section~\ref{S_s:motivation_spread} for an example of this phenomenon. An encompassing GFT framework was proposed in~\cite{girault2018gft}, which combines aspects of signal variation and energy to design general orthonormal transforms for graph signals.

\noindent\textbf{Contributions and paper outline.} Here we design a novel DGFT that has the following desirable properties: P1) The \blue{basis graph signals} provide notions of frequency and signal variation over digraphs which are also consistent with {those typically used for} subsumed undirected graphs. P2) Frequency components are designed to be (approximately) equidistributed in $[0,f_{\max}]$, and thus better capture low, middle, and high frequencies. P3) \blue{Basis vectors} are orthonormal so Parseval's identity holds and inner products are preserved in the vertex and graph frequency domains. Moreover, the inverse DGFT can be easily computed. To formalize our design goal via a well-defined criterion, in Section~\ref{S:Notation} we introduce  a smooth spectral dispersion function, which measures the spread of the frequencies associated with candidate DGFT  \blue{basis vectors}. Motivation and challenges associated with the proposed spectral-dispersion minimization approach are discussed in Section~\ref{S:spread_max_freq}. We then propose two algorithmic approaches with complementary strengths, to construct a DGFT basis with the aforementioned properties P1)-P3). We first leverage a provably-convergent feasible method for orthonormality-constrained optimization, to directly minimize the smooth spectral-dispersion cost over the Stiefel manifold (Section \ref{S:feasible_method}). In Section \ref{S:heuristic}, we propose a DGFT heuristic whereby we restrict the set of candidate frequency components to the (possibly sign reversed) eigenvectors of the Laplacian matrix associated with an undirected version of $\ccalG$. In this setting, we show that the spectral-dispersion minimization problem can be cast as supermodular (frequency) set optimization. Moreover, we show that orthonormality can be enforced via a matroid basis constraint, which motivates the adoption of a scalable greedy algorithm to obtain an approximate solution with provable worst-case performance guarantees. 

 \blue{Relative to the conference precursors to this paper~\cite{shafipour2017digraph,shafipour2018digraph}, here we offer a more comprehensive and unified exposition of the aforementioned algorithmic approaches, comparing their computational complexity and their performance against state-of-the-art GFT methods for digraphs (Section~\ref{S:numerical}). This way, results become clearer and new insights as well as conclusions can be drawn. We include all proofs and technical details missing from~\cite{shafipour2017digraph,shafipour2018digraph}, as well as new examples and results that shed light on the properties of directed variation, the maximum attainable frequency on a digraph, and the proposed dispersion minimization problem.} Numerical tests with both synthetic and real-world digraphs corroborate the effectiveness of our DGFT algorithms in yielding (near) maximally-spread frequency components (Section~\ref{S:numerical}). Concluding remarks  \blue{and future research directions are outlined} in Section~\ref{S:Conclusions}, while some technical details are deferred to the Appendix.

\noindent\textbf{Notation.} Bold capital letters refer to matrices and bold lowercase letters represent vectors. The entries of a matrix $\mathbf{X}$ and a (column) vector $\mathbf{x}$ are denoted by $X_{ij}$ and $x_i$, respectively. Sets are represented by calligraphic capital letters. 
The notation $^T$ stands for transposition and $\bbI_N$ represents the $N\times N$ identity matrix, while $\mathbf{1}_N$ denotes the $N\times 1$ vector of all ones. For a vector $\bbx$, $\textrm{diag}(\bbx)$ is a diagonal matrix whose $i$th diagonal entry is $x_i$.  \blue{The binary-valued indicator function of a statement $S$ is denoted by $\ind{S}$, where $\ind{S}=1$ if $S$ holds true and $\ind{S}=0$ otherwise.} Lastly, $\| \bbx \|=(\sum_{i} x_{i}^{2})^{1/2}$ denotes the Euclidean norm  of $\bbx$, while $\textrm{trace}(\bbX)=\sum_{i} X_{ii}$ stands for the trace of a square matrix $\bbX$ whose Frobenius norm is  \blue{$\|\bbX\|_F=\big[\textrm{trace}(\bbX\bbX^T)\big]^{1/2}$}. 
\section{Preliminaries and Problem Statement}\label{S:Notation}
In this section we extend the notion of signal variation to digraphs and accordingly define graph frequencies. We then state the problem of finding an orthonormal DGFT basis with evenly distributed frequencies in the graph spectral domain.

{\subsection{Signal variation on digraphs} \label{S:s_signal_variation}
	 Our goal is to find $N$ orthonormal  \blue{vectors} capturing {different} modes of variation over the graph $\ccalG$. We collect these desired  \blue{vectors} in a matrix $\bbU \coloneqq [\bbu_1, \ldots, \bbu_N] \in \reals^{N \times N}$, where $\bbu_k \in \reals^N$ represents the $k$th frequency component. This means that the DGFT of  a graph signal $\bbx\in\reals^N$ with respect to $\ccalG$ is the signal $\tbx=[\tdx_1,\ldots,\tdx_{N}]^T$ defined as $\tbx = \bbU^T\bbx$. The inverse (i)DGFT of $\tbx$ is given by $\bbx = \bbU \tbx=\sum_{k=1}^{N} \tdx_k \bbu_k$, which allows one to synthesize $\bbx$ as a sum of orthogonal frequency components $\bbu_k$. The contribution of $\bbu_k$ to the signal $\bbx$ is the DGFT coefficient $\tdx_k$.
	 
	 For undirected graphs, the quantity $\text{TV}(\bbx)$ in \eqref{e:TV_def} measures how signal $\bbx$ varies over the {network} with Laplacian $\bbL$. This motivates defining a more general notion of signal variation for digraphs, called directed variation (DV), as
	\begin{equation} \label{e:DV_def}
		\text{DV}(\bbx) \coloneqq \sum_{i,j=1}^{N} A_{ij} [x_i - x_j]_{+}^2,
	\end{equation}
	where $[x]_{+} \coloneqq \max(0,x)$ denotes projection onto the non-negative reals.  \blue{Notice that $[x]_+^2=([x]_+)^2$, and for brevity we omit the parenthesis. Similar to the total variation \eqref{e:TV_def}, the proposed directed variation measure is non-negative, non-linear, but convex and differentiable.} To gain insights on \eqref{e:DV_def}, consider a graph signal $\bbx \in \reals^N$ on digraph $\ccalG$ {and suppose a} directed edge represents the direction of signal flow from a larger value to a smaller one. Thus, an edge from node $i$ to node $j$ {(i.e., $A_{ij}>0$) contributes to $\text{DV}(\bbx)$ only if $x_i > x_j$. Accordingly, one in general has that $\text{DV}(\bbx)\neq \text{DV}(-\bbx)$ and we will exploit this property later.
	Notice that if $\ccalG$ is undirected, then $\text{DV}(\bbx)\equiv\text{TV}(\bbx)$  \blue{because from $A_{ij} [x_i - x_j]_{+}^2$ and $A_{ji} [x_j - x_i]_{+}^2$, one will be zero and the other one will be equal to $A_{ij} (x_i - x_j)^2$}.} 

Analogously to the undirected case, we define the frequency $f_k := \text{DV}(\bbu_k)$ as the directed variation of the  \blue{vector} $\bbu_k$.
	
	{\subsection{Challenges facing spread frequency components} \label{S:s_challenges}
		Similar to the discrete spectrum of periodic time-varying signals, by designing the \blue{basis signals} we would ideally like to have} $N$ equidistributed {graph} frequencies forming an arithmetic sequence
	\begin{equation} \label{e:directed_freq}
		f_k = {\text{DV}(\bbu_k)=} \frac{k-1}{N-1} f_{\max}, \quad k=1,\ldots,N
	\end{equation}
	where $f_{\max}$ is the maximum variation attainable by a unit-norm signal on $\ccalG$. Such a spread frequency distribution could facilitate more interpretable spectral analyses of graph signals (where it is apparent what low, medium and high frequencies mean), and also aid filter design in the graph spectral domain.
	
	Not surprisingly, finding a DGFT basis attaining the exact frequencies in \eqref{e:directed_freq} may be impossible for irregular graph domains. This can be clearly seen for undirected graphs, where one has the additional constraint that the summation of frequencies is constant, since
	\begin{equation}
	\sum_{k=1}^{N}f_k  =  \sum_{k=1}^{N}\text{TV}(\bbu_k)=\textrm{trace}(\bbU^{T}\bbL\bbU)=\textrm{trace}(\bbL).
	\end{equation}
	
	{Moreover}, one needs to determine the maximum frequency $f_{\max}$ that a unit-norm graph signal can attain. For undirected graphs, one has
	\begin{equation} \label{e:lambda_max}
		f_{\max}^{u} := \max_{\lVert \bbu \rVert = 1} \text{TV}(\bbu) =\max_{\lVert \bbu \rVert = 1} \bbu^T \bbL \bbu = \lambda_{\max},
	\end{equation}
	where $\lambda_{\max}$ is the largest eigenvalue of the Laplacian matrix $\bbL$. However, finding the maximum \emph{directed variation} is in general challenging, since one needs to solve {the (non-convex) spherically-constrained problem}
	\begin{equation}\label{e:opt_f_max}
		\bbu_{\max} = \argmax_{\lVert \bbu \rVert = 1} \: \text{DV}(\bbu) \quad \text{and} \quad f_{\max} := \text{DV}(\bbu_{\max}).
	\end{equation}
	To relate the maximum frequencies in \eqref{e:lambda_max} and \eqref{e:opt_f_max}, for a given digraph $\ccalG=(\ccalV,\bbA)$ consider its \emph{underlying undirected graph} $\ccalG^u=(\ccalV,\bbA^u)$, obtained by replacing all directed edges in $\ccalG$ with undirected ones via $A^u_{ij} = {A^u_{ji}}:=\max(A_{ij}, A_{ji})$. Notice then that $f_{\max}$ is upper-bounded by $f_{\max}^{u}=\lambda_{\max}$, the spectral radius of the Laplacian of $\ccalG^u$.  \blue{This is formally proved in Proposition~\ref{pro_lambda_upper}, but it is essentially because dropping the direction of any edge can not decrease the directed variation of a signal}.
	
	\subsection{Problem statement} \label{S:s_problem}
	
	Going back to the design of $\bbU$, to cover the whole spectrum of variations one would like to set $\bbu_1=\bbu_{\min}:=\frac{1}{\sqrt{N}}\mathbf{1}_{N}$ (normalized all ones vector of length $N$\blue{,} i.e., a constant signal) and $\bbu_N=\bbu_{\max}$ in \eqref{e:opt_f_max}. As a criterion for the design of the remaining  \blue{basis vectors}, consider the spectral dispersion function
		\begin{equation} \label{e:dispersion_def}
			\delta(\bbU):=\sum_{i=1}^{N-1} \left[\text{DV}(\bbu_{i+1})-\text{DV}(\bbu_i)\right]^2
		\end{equation}
		that measures how well spread the corresponding frequencies are over $[0,f_{\max}]$.   Having fixed the first and last columns of $\bbU$, it follows that the dispersion function $\delta(\bbU)$ is minimized when the free directed variation values are selected to form an arithmetic sequence between $\text{DV}(\bbu_1)=0$ and $\text{DV}(\bbu_N)=f_{\max}$, consistent with our design goal. 
		
Rather than going after frequencies exactly equidistributed as in \eqref{e:directed_freq}, our idea is to minimize the spectral dispersion 
\begin{align}
\label{eq:delta_opt_prob}
& \min_{\bbU}
& & \hspace{-2cm}\sum_{i=1}^{N-1} \left[\text{DV}(\bbu_{i+1})-\text{DV}(\bbu_i)\right]^2 \\
& \text{subject to}
& & \hspace{-1.5cm} \bbU^T\bbU=\bbI_N,\nonumber\\
&
&&\hspace{-1.5cm} \bbu_1=\bbu_{\min},\nonumber\\
&
&& \hspace{-1.5cm}\bbu_N=\bbu_{\max}.\nonumber
\end{align}
Problem \eqref{eq:delta_opt_prob} is feasible since we show in Appendix \ref{S:s_feasibility} that $\bbu_{\max}$ defined in \eqref{e:opt_f_max} is orthogonal to the constant vector $\bbu_{\min}=\frac{1}{\sqrt{N}}\mathbf{1}_{N}$.  \blue{There is no need to explicitly add a constraint enforcing that basis vectors $\bbu_i$ are ordered, meaning that their respective $\textrm{DV}(\bbu_i)$ values  increase with $i$.  The solution of \eqref{eq:delta_opt_prob} will be ordered, otherwise one could simply sort the candidate directed variation values and decrease the objective. As an alternative definition of graph signal energy, one could adopt generalized orthonormality constraints $\bbU^T\bbM\bbU=\bbI_N$ for some positive definite $\bbM$~\cite{girault2018gft}.} In any case, finding the global optimum of \eqref{eq:delta_opt_prob} is challenging due to the  non-convexity arising from the orthonormality (Stiefel manifold) constraints  \blue{as well as the objective function [due to the cross-terms $\text{DV}(\bbu_i)\text{DV}(\bbu_{i+1})$].} The spectral dispersion $\delta (\bbU)$ is smooth though, and so there is hope of finding good stationary solutions by bringing to bear recent advances in manifold optimization.

In Section \ref{S:feasible_method} we build on a feasible method for optimization with orthogonality constraints~\cite{wen2013feasible}, to solve judiciously modified forms of problems \eqref{e:opt_f_max} and \eqref{eq:delta_opt_prob} to directly find the maximum frequency along with the disperse basis vectors. 
But before delving into algorithmic solutions, in the next section we expand on the motivation behind spectral dispersion minimization. 
We also offer additional graph-theoretic insights on the maximum directed variation a unit norm signal can achieve.

\section{On Spread and Maximum Digraph Frequencies}\label{S:spread_max_freq}

Here, we further motivate the advocated DGFT design by first showing how state-of-the-art methods may fail to offer signal representations capturing different modes of signal variation with respect to $\ccalG$. For undirected graphs where the Laplacian eigenbasis has well documented merits, we then show that the said GFT in general does not minimize the spectral dispersion measure \eqref{e:dispersion_def}. In Section~\ref{S:s_freq_max} we revisit problem~\eqref{e:opt_f_max}, namely that of finding the maximum frequency over a given digraph. We first identify some graph families for which the maximum directed variation can be obtained analytically, and then provide a general 1/2-approximation to $f_{\max}$ that will serve  as a first step for a
greedy DGFT construction algorithm in Section~\ref{S:heuristic}.

\subsection{Motivation for spread frequencies} \label{S_s:motivation_spread}

\begin{figure}[t]
	\centering    
	{\includegraphics[width=0.8\linewidth]{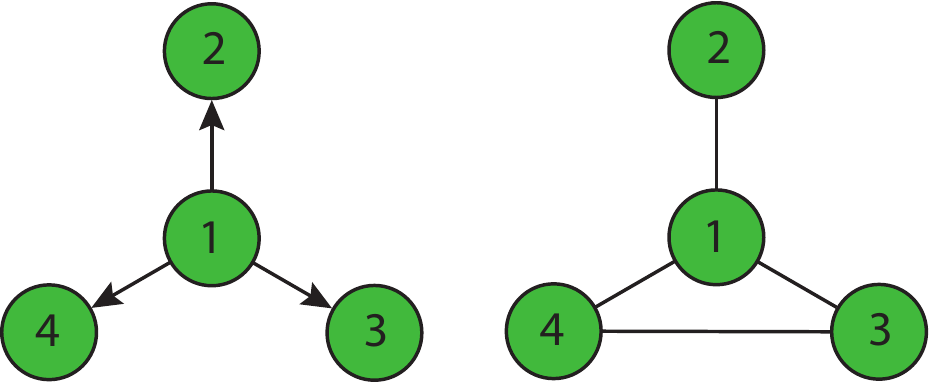}}
	\caption{Toy directed (left) and undirected (right) graphs used to motivate the advocated DGFT design based on spectral dispersion minimization. For the shown digraph, the GFT construction of \cite{sardellitti} fails to yield signal expansions with respect to different modes of variation. The Laplacian eigenbasis of the shown undirected graph does not minimize the dispersion criterion in \eqref{eq:delta_opt_prob}.}
	\label{fig:toy}
		\vspace{-0.4cm}
\end{figure}

\noindent\textbf{Directed graphs.} The directed variation measure
\begin{equation}\label{e:dv_sardellitti}
\mathrm{DV}'(\bbx) \coloneqq \sum_{i,j=1}^{N} A_{ij} [\bbx_i - \bbx_j]_{+}
\end{equation}
was introduced in \cite{sardellitti} as the convex Lov\'asz extension of the graph cut size, whose minimization can facilitate identifying graph clusters~\cite[Ch. 4]{kolaczyk2009statistical}. Different from \eqref{e:DV_def}, the directed variation measure \eqref{e:dv_sardellitti} is not smooth and for undirected graphs it boils down to a so-termed graph absolute variation $\text{TV}_1(\bbx):=\sum_{i,j=1,j>i}^{N} A_{ij} |x_i - x_j|$ [cf. \eqref{e:TV_def}]; see also \cite{DSP_freq_analysis,chen_variation_min}.  To obtain a GFT for digraphs, the approach in~\cite{sardellitti} is to solve
the orthogonality-constrained problem 
%
\begin{equation}
\label{eq:sardellitti_opt_prob}
\min_{\bbU} \sum_{i=1}^{N} \mathrm{DV}'(\bbu_{i}),\quad
\text{subject to }\bbU^T\bbU=\bbI_N.
\end{equation}
An attractive feature of this construction is that it can offer parsimonious representations of graph signals exhibiting smooth structure within clusters (i.e., densely connected subgraphs) of the underlying graph $\ccalG$.

Consider {the} digraph with $N=4$ nodes shown in Fig. \ref{fig:toy} (left). {An optimal GFT basis $\bbU$} solving \eqref{eq:sardellitti_opt_prob} takes the form
\begin{equation*}
\bbU=
\begin{bmatrix}
0.5 & c  & c & c\\
0.5 & a  & 0 & b\\
0.5 & b  & a & 0\\
0.5 & 0  & b & a
\end{bmatrix},
\end{equation*}
where $a=(1+\sqrt{5})/4\approx 0.8090$, $b=(1-\sqrt{5})/4\approx -0.3090$, and $c=-0.5$. These values satisfy $a+b+c=0$, $a^2+b^2+c^2=1$, and $c^2+ab=0$, which implies the orthonormality of $\bbU$. {As a result,} for all columns $\bbu_k$ of $\bbU$ one has $\mathrm{DV}'(\bbu_k)=0,\: k=1,\ldots, 4,$ and hence the inverse GFT {synthesis} formula $\bbx = \bbU \tilde{\bbx}$ {fails to} offer an expansion of {\bbx} with respect to \emph{different} modes of variation {(e.g., low and high graph frequencies)}.

\begin{figure}[t]
	\centering   
	{\includegraphics[width=1\linewidth]{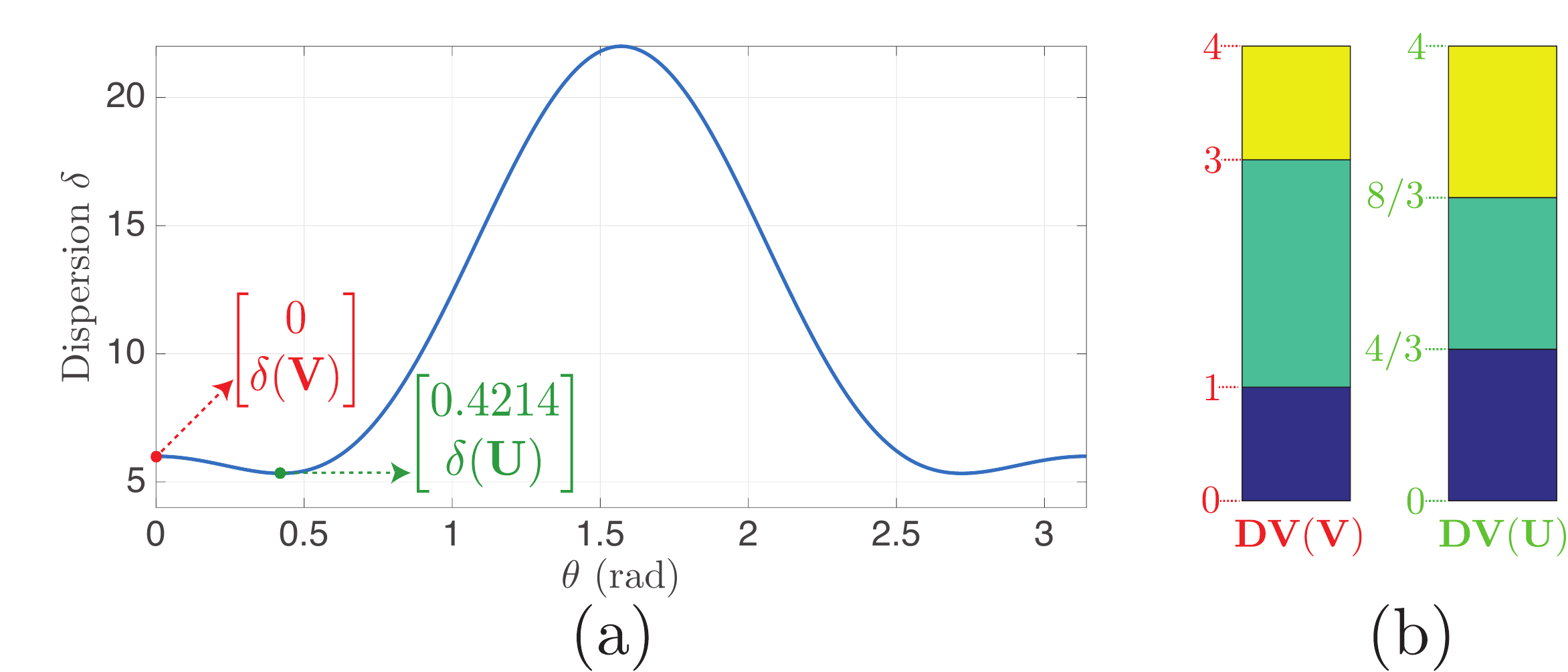}}
	\vspace{-0.7cm}
	\caption{(a) Spectral dispersion function  \blue{\eqref{e:dispersion_def}} versus $\theta$ in radians, for the undirected graph in Fig.~\ref{fig:toy} (right). Apparently, the solution $\bbU$ of \eqref{eq:delta_opt_prob} offers a more disperse basis than the eigenvectors $\bbV$ of the Laplacian matrix. (b) Colored boxes show the consecutive frequency differences obtained by $\bbV$ (left) and $\bbU$ (right), while the specific directed variation values correspond to the horizontal boundary lines. In this particular case, the optimum basis $\bbU$ yields exactly equidistributed graph frequencies in $[0,4]$, as defined in \eqref{e:directed_freq}.}
	\label{fig:undirected_dispersion}
		\vspace{-0.4cm}
\end{figure}


\noindent\textbf{Undirected graphs.} When the edges of the graph are undirected, the workhorse GFT approach is to project the signals onto the eigenvectors of the graph Laplacian $\bbL$; see e.g.,~\cite{gsp2018tutorial,tremblay2017design,zhu2012approximating}. A pertinent question is whether the Laplacian-based GFT minimizes the spectral dispersion function in \eqref{e:dispersion_def}, where $\text{DV}(\bbx)\equiv\text{TV}(\bbx)$ because $\ccalG$ is undirected.  To provide an answer, we consider the graph shown in Fig.~\ref{fig:toy} (right)  \blue{and form its Laplacian matrix $\bbL$}. Denote the eigenvectors of $\bbL$ as $\bbV=[\bbv_1,\bbv_2,\bbv_3,\bbv_4]$, with corresponding frequencies $[\lambda_1, \lambda_2, \lambda_3, \lambda_4] = [0, 1, 3, 4]$. To determine the DGFT basis $\bbU$ satisfying our design criteria in \eqref{eq:delta_opt_prob}, we first set $\bbu_1 := \bbv_1$ and $\bbu_4 := \bbv_4$ which represent the frequency components associated with minimum and maximum frequencies, respectively. Since the Laplacian eigenvalues are distinct and we are searching for an orthonormal basis, then it follows that $\bbv_2$ and $\bbv_3$ span the hyperplane containing $\bbu_2$ and $\bbu_3$. There is a single degree of freedom to specify $\bbu_2$ and $\bbu_3$ on that plane, namely a simultaneous rotation of $\bbv_2$ and $\bbv_3$. Accordingly, all the feasible basis vectors $\{\bbu_2,\bbu_3\}$ will be of the form
\begin{equation*} 
 \blue{
\begin{bmatrix}
\bbu_2 & \bbu_3 
\end{bmatrix}
=
\begin{bmatrix}
\bbv_2  & \bbv_3 
\end{bmatrix}
\bbR_\theta
=
\begin{bmatrix}
\bbv_2  & \bbv_3 
\end{bmatrix}
\begin{bmatrix}
\cos \theta &  \sin\theta \\
-\sin \theta &   \cos \theta \\
\end{bmatrix}
}
,
\end{equation*}
where $\bbR_\theta$ rotates vectors counterclockwise by an angular amount of $\theta$. Collecting the sought  \blue{basis vectors} in $\bbU_{\theta}=[\bbv_1,\bbu_2,\bbu_3,\bbv_4]$, in Fig.~\ref{fig:undirected_dispersion} (a) we plot the dispersion function  \blue{defined in \eqref{e:dispersion_def}}
as a function of $\theta$. 
It is apparent from Fig.~\ref{fig:undirected_dispersion}-(a) that $\delta(\bbV)$ obtained from eigenvectors of the Laplacian is not a global minimizer of the spectral dispersion function. As shown in Fig.~\ref{fig:undirected_dispersion}-(a), the optimum basis is $\bbU := \bbU_\theta|_{\theta\approx 0.4214 }$. 

To further compare the frequency components obtained, the horizontal lines in Fig~\ref{fig:undirected_dispersion}-(b) depict both sets of frequencies $\{\text{DV}(\bbv_k)\}_{k=1}^4$ and $\{\text{DV}(\bbu_k)\}_{k=1}^4$. As expected, the optimized GFT basis $\bbU$ gives rise to frequencies that are more uniformly spread in the graph spectral domain. Moreover, for this particular example the graph frequencies $\{\text{DV}(\bbu_k)\}_{k=1}^4$ form an arithmetic sequence in $[0,4]$, which for general graphs may be infeasible as discussed in Section~\ref{S:s_challenges}. In Sections~\ref{S:feasible_method} and \ref{S:heuristic} we propose two approaches with complementary strengths to find spread frequencies for arbitrary digraphs. 

\subsection{Maximum directed variation} \label{S:s_freq_max}
As mentioned in {Section} \ref{S:s_challenges}, one challenge in finding an approximately equidistributed set of frequencies on a digraph $\ccalG(\ccalV,\bbA)$ is to calculate the maximum frequency $f_{\max}$.
The spherically-constrained problem \eqref{e:opt_f_max} is non-convex and in general challenging to solve  [cf.~\eqref{e:lambda_max} for subsumed undirected graphs, whose solution is the spectral radius of the Laplacian]. The following proposition asserts that for some particular classes of digraphs (depicted in Fig~\ref{fig:f_max_special}), the value of $f_{\max}$ can be obtained analytically. 

\begin{myproposition}
	\normalfont
\label{pro:special_cases}
Let $f_{\max}$ be the maximum directed variation that a unit norm vector can attain as defined in \eqref{e:opt_f_max}.
\begin{enumerate}
\item Let $\ccalG$ be the \emph{directed path (dipath)} depicted in Fig.~\ref{fig:f_max_special}-(a), i.e., a digraph whose adjacency matrix has nonzero entries $A_{ij}>0$ only for $j=i+1$, $i<N$. Then, $f_{\max}=2 \max_{i,j} A_{ij}$.
\item Let $\ccalG$ be the \emph{directed cycle} depicted in Fig.~\ref{fig:f_max_special}-(b), i.e., a digraph whose adjacency matrix has nonzero entries $A_{ij}>0$ only for $j=\textrm{mod}_N(i)+1$, where $\textrm{mod}_N(x)$ denotes the modulus (remainder) obtained after dividing $x$ by $N$. Then, $f_{\max}=2 \max_{i,j} A_{ij}$.
\item Let $\ccalG$ be a \emph{unidirectional bipartite graph} as depicted in Fig.~\ref{fig:f_max_special}-(c), i.e., a bipartite digraph where $\ccalV=\ccalV^+\cup\ccalV^-$, $\ccalV^+\cap\ccalV^-=\emptyset$, and whose adjacency matrix may only have nonzero entries $A_{ij}>0$ for $i\in\ccalV^+$ and $j\in\ccalV^-$. Let $\bbL$ be the Laplacian matrix of the underlying undirected graph $\ccalG^{u}$ [recall the discussion following \eqref{e:opt_f_max}], with spectral radius $\lambda_{\max}$. Then, $f_{\max}=\lambda_{\max}$.
\end{enumerate}
\end{myproposition}
\begin{myproof}
See the Appendices \ref{S:s_path_proof}, \ref{S:s_cycle_proof} and \ref{S:s_bipartite_proof}.
\end{myproof}

Beyond the special digraph families in Proposition~\ref{pro:special_cases}, it is not clear how to solve \eqref{e:opt_f_max}.  \blue{Next we show that $f_{\max}$ is upper-bounded by $\lambda_{\max}$, the spectral radius of the Laplacian of the underlying undirected graph $\ccalG^u$; see Section~\ref{S:s_challenges}.}

\begin{figure}[t]
	\centering    
		{\includegraphics[width=0.6\linewidth]{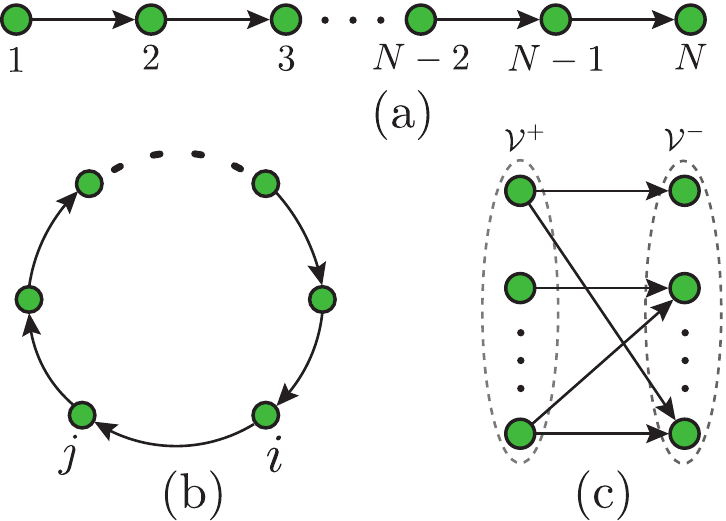}}
	\caption{Some special families of digraphs for which $f_{\text{max}}$ as defined in \eqref{e:opt_f_max} can be obtained analytically. (a)~Directed path; (b)~directed cycle; and (c)~unidirectional bipartite graph, where all the edges go from vertices in $\ccalV^+$ to those in $\ccalV^-$.}
	\label{fig:f_max_special}
	\vspace{-0.4cm}
\end{figure}

 \blue{
\begin{myproposition}
		 \normalfont
		\label{pro_lambda_upper}
For a digraph $\ccalG$, recall its underlying undirected graph $\ccalG^{u}$ and the spectral radius $\lambda_{\mathrm{max}}$ of its Laplacian $\bbL$. Then, the maximum directed variation $f_{\max}$ defined in \eqref{e:opt_f_max} is upper-bounded by $f_{\max}^{u}=\lambda_{\max}$.
    \end{myproposition}
\begin{myproof}
From the definition of directed variation \eqref{e:opt_f_max}, we have
\begin{equation}
\begin{aligned}
 f_{\text{max}}  & = \sum_{i,j=1}^{N} A_{ij} [u_{\max,i} - u_{\max,j}]_{+}^2 \\ \nonumber
& = \sum_{\substack{i,j=1 \\ i<j}}^{N} A_{ij} [u_{\max,i} - u_{\max,j}]_{+}^2 + A_{ji} [u_{\max,j} - u_{\max,i}]_{+}^2.
\end{aligned}
\end{equation}
Since $A_{ij}, A_{ji} \leq \max\{A_{ij},A_{ji}\} = A_{ij}^{u}$, then it follows that
\begin{equation}
\begin{aligned}
 f_{\text{max}}  & \leq\sum_{\substack{i,j=1 \\ i<j}}^{N} A_{ij}^{u} \bigg( [u_{\max,i} - u_{\max,j}]_{+}^2 + [u_{\max,j} - u_{\max,i}]_{+}^2 \bigg) \\ \nonumber
 & = \sum_{i,j=1, i<j}^{N} A_{ij}^{u} \big(u_{\max,i} - u_{\max,j}\big)^{2} = \text{TV}(\bbu_{\max}).
\end{aligned}
\end{equation}
Finally, because $\text{TV}(\bbu_{\max})$ is upper-bounded by $\lambda_{\max}$ in \eqref{e:lambda_max}, we conclude that $ f_{\text{max}}  \leq \lambda_{\max}$.
\end{myproof}
}

For general digraphs, Proposition~\ref{pro:f_approx} specifies how to find a basis vector $\tilde{\bbu}$ with an approximate $\tilde{f}_{\max}\coloneqq \text{DV}(\tilde{\bbu})$ which is at least half of $f_{\max}$. Once more, the underlying undirected graph $\ccalG^{u}$ and the leading eigenvector of its Laplacian matrix will prove instrumental to obtain the desired approximation.
\begin{myproposition}
	\normalfont
	\label{pro:f_approx}
	For a digraph $\ccalG$, {recall its underlying undirected graph $\ccalG^{u}$ and the spectral radius $\lambda_{\mathrm{max}}$ of its Laplacian $\bbL$. Let $\bbu$ be the dominant eigenvector of $\bbL$, i.e., the unit-norm vector $\bbu$ such that $\bbL\bbu=\lambda_{\max}\bbu$.} Then,  \blue{ a worst-case $1/2$-approximation to $f_{\max}$ is given by}
	\begin{equation}
	\label{eq:f_tilde}
		\tilde{f}_{\max}\coloneqq \max{\{\mathrm{DV}(\bbu),\mathrm{DV}(-\bbu)\}}\geq \frac{f_{\max}}{2}.
	\end{equation} 
\end{myproposition}
\begin{myproof}
	First recall that
	%
	\begin{equation*} 
	\lambda_{\max}  =  \bbu^T\bbL\bbu = \sum_{\substack{i,j=1 \\ i<j}}^{N} A^u_{ij} (u_i - u_j)^2 = \frac{1}{2}\sum_{i,j=1}^{N} A^u_{ij} (u_i - u_j)^2.
	\end{equation*}
	Since $A^u_{ij}\leq A_{ij}+A_{ji}$ and $(u_i - u_j)^2=[u_i - u_j]_{+}^2+[u_j - u_i]_{+}^2$, 
	\begin{align*} 
		\lambda_{\text{max}}\leq {}& \frac{1}{2}\sum_{i,j=1}^{N} (A_{ij}+A_{ji}) \bigg([u_i - u_j]_{+}^2+[u_j - u_i]_{+}^2\bigg)\nonumber \\
		={} & \text{DV}(\bbu)+\text{DV}(-\bbu).
	\end{align*}
	{In conclusion}, at least one of $\text{DV}(\bbu)$ or $\text{DV}(-\bbu)$ is larger than $\lambda_{\text{max}}/2$, and this completes the proof since $\lambda_{\text{max}}\geq f_{\text{max}}$.
\end{myproof}

\noindent In practice, we can {compute} $\max{\{\text{DV}(\bbu),\text{DV}(-\bbu)\}}$  {for} \emph{any}  eigenvector {$\bbu$} of the Laplacian matrix. This will possibly give a higher frequency {in $\ccalG$}, while preserving the $1/2$-approximation. 

In Section~\ref{S:heuristic} we will revisit the result in Proposition~\ref{pro:f_approx}, to motivate a greedy heuristic to construct a disperse DGFT  basis from Laplacian eigenvectors of $\ccalG^u$. But before that, in the next section we develop a DGFT algorithm that in practice returns near-optimal solutions to the spectral dispersion minimization problem~\eqref{e:dispersion_def}. While computationally more demanding than the recipe in Proposition~\ref{pro:f_approx}, we show \blue{in Section~\ref{Ss:max_freq_feasible} that} the adopted framework for orthogonality-constrained optimization can be as well used to accurately approximate $f_{\max}$. 
\section{Minimizing Dispersion in a Stiefel Manifold}\label{S:feasible_method}
Here we show how to find a disperse Fourier basis for signals on digraphs, by bringing to bear a feasible method for optimization of differentiable functions over the Stiefel manifold \cite{wen2013feasible}. Specifically, following the specification in Section \ref{S:s_problem} we take a two step approach whereby: i) we find $f_{\max}$ and its corresponding basis \blue{vector} $\bbu_{\max}$ by solving \eqref{e:opt_f_max}; and ii) we solve \eqref{eq:delta_opt_prob} to find well-spread frequency components $\bbU=[\bbu_1, \cdots, \bbu_N]$ satisfying $\bbu_1 = \bbu_{\min}=\frac{1}{\sqrt{N}}\mathbf{1}_N$ and $\bbu_N = \bbu_{\max}$. Similar feasible methods have been also successfully applied to a wide variety of applications, such as low-rank matrix approximations, Independent Component Analysis,  and subspace tracking, to name a few~\cite{absil2009optimization}.

The general iterative method of \cite{wen2013feasible} deals with an orthogonality constrained problem of the form
\begin{equation} \label{e:general_feasible_orthogonal}
\min _{\bbU \in \reals^{n \times p}} \phi(\bbU), \quad \text{subject to} \quad \bbU^{T} \bbU = \bbI_p,
\end{equation}
where $\phi (\bbU): \reals^{n \times p} \rightarrow \reals$ is assumed to be differentiable, just like $\delta(\bbU)$ in \eqref{e:dispersion_def}. Given a feasible point $\bbU_{k}$ at iteration $k=0,1,2,\ldots$ and the gradient $\bbG_{k} = \nabla \phi (\bbU_{k})$, one follows the update rule
\begin{equation} \label{e:feasible_update_rule}
\bbU_{k+1} (\tau) = \left(\bbI_n+\frac{\tau}{2} \bbB_{k}\right)^{-1}\left(\bbI_n - \frac{\tau}{2} \bbB_{k}\right)\bbU_{k},
\end{equation}
where $\bbB_{k} := \bbG_{k} \bbU_{k}^{T} - \bbU_{k} \bbG_{k}^{T}$ is a skew-symmetric ($\bbB_{k}^{T} = -\bbB_{k}$)  projection of the gradient onto the constraint's tangent space. Update rule \eqref{e:feasible_update_rule} is known as the Cayley transform which preserves orthogonality (i.e., $\bbU_{k+1}^{T} \bbU_{k+1} = \bbI_p$), since $(\bbI_n+\frac{\tau}{2} \bbB_{k})^{-1}$ and $\bbI_n - \frac{\tau}{2} \bbB_{k}$ commute. Other noteworthy properties of the update are: i) $\bbU_{k+1}(0) = \bbU_{k}$; ii) $\bbU_{k+1}(\tau)$ in \eqref{e:feasible_update_rule} is smooth as a function of the step size $\tau$; and iii) $\frac{d}{d\tau}\bbU_{k+1}(0)$ is the projection of $-\bbG_{k}$ into the tangent space of the Stiefel manifold at $\bbU_{k}$. 

Most importantly, iii) ensures that the update \eqref{e:feasible_update_rule} is a descent path for a proper step size $\tau$, which can be obtained through a curvilinear search satisfying the Armijo-Wolfe conditions
\begin{subequations}
\begin{align} 
\phi(\bbU_{k+1}(\tau_{k})) &\leq \phi(\bbU_{k+1}(0)) + \rho_{1} \tau_{k} \phi'_{\tau}(\bbU_{k+1}(0))\label{e:Armijo-Wolfe_conditionsa} \\
\phi'_{\tau}(\bbU_{k+1}(\tau_{k})) &\geq \rho_{2} \phi'_{\tau}(\bbU_{k+1}(0))\label{e:Armijo-Wolfe_conditionsb},
\end{align}
\end{subequations}
where $0<\rho_{1}<\rho_{2}<1$ are two parameters \cite{jorge2006numerical}.
One can show that if $\phi(\bbU_{k+1}(\tau))$ is continuously differentiable and bounded below as is the case for problems \eqref{e:opt_f_max} and \eqref{eq:delta_opt_prob}, then there exists a $\tau_{k}$ satisfying \eqref{e:Armijo-Wolfe_conditionsa} and \eqref{e:Armijo-Wolfe_conditionsb}. Moreover, the derivative of $\phi(\bbU_{k+1}(\tau))$ at $\tau = 0$ is given by $\phi'_{\tau}(\bbU_{k+1}(0)) = -1/2 \lVert \bbB_{k} \rVert_{F}$; see~\cite{wen2013feasible} for additional details. All in all, the iterations~\eqref{e:feasible_update_rule} are well defined and by implementing the aforementioned curvilinear search, \cite[Theorem 2]{wen2013feasible} asserts that the overall procedure converges to a stationary point of $\phi(\bbU)$, while generating feasible points in the Stiefel manifold at every iteration.


\begin{algorithm}[t]
	\caption{Directed Variation Maximization}
	\label{alg:feasible_f_max}
	\begin{algorithmic}[1]
		\STATE \textbf{Input:} Adjacency matrix $\bbA$ and parameter $\epsilon > 0$.
		\STATE \textbf{Initialize} $k=0$ and unit-norm $\bbu_{0} \in \reals^{N}$ at random.
		\REPEAT
		\STATE Evaluate objective $\phi(\bbu_{k}):=-\text{DV}(\bbu_{k})$ in \eqref{e:DV_def}.
		\STATE Compute gradient $\bar{\bbg}_{k} \in \reals^{N}$ via \eqref{e:f_max_gradient}.
		\STATE Form $\bbB_{k} = \bar{\bbg}_{k} \bbu_{k}^{T} - \bbu_{k} \bar{\bbg}_{k}^{T}$.
		\STATE Select $\tau_{k}$ satisfying conditions \eqref{e:Armijo-Wolfe_conditionsa} and \eqref{e:Armijo-Wolfe_conditionsb}.
		\STATE Update $\bbu_{k+1} (\tau_{k}) = (\bbI_N+\frac{\tau_{k}}{2} \bbB_{k})^{-1}(\bbI_N - \frac{\tau_{k}}{2} \bbB_{k})\bbu_{k}$.
		\STATE $k \gets k+1$.
		\UNTIL{$\lVert \bbu_{k}- \bbu_{k-1} \rVert \leq \epsilon$}
		\STATE \textbf{Return} $\bbu_{\max}:=\bbu_{k}$ and $f_{\max}:=\text{DV}(\bbu_{\max})$.
	\end{algorithmic}
\end{algorithm}

\subsection{Directed variation maximization} \label{Ss:max_freq_feasible}
As the first step to find the DGFT \blue{basis}, we obtain $f_{\max}$ by using the feasible approach to minimize $-$DV($\bbu$) over the sphere $\{ \bbu\in\reals^N \mid \bbu^T \bbu = 1  \}$ [cf. \eqref{e:opt_f_max}]. The gradient $\bar{\bbg} := \nabla\text{DV}(\bbu) \in \reals^N$ has entries $\bar{g}_i,\: 1 \leq i \leq  N,$ given by 
\begin{equation} \label{e:f_max_gradient}
\bar{g}_i = 2\left(\bbA_{\cdot i} ^{T}[\bbu-u_i \mathbf{1}_{N}]_{+}  - \bbA_{i\cdot}[u_i \mathbf{1}_{N}-\bbu]_{+} \right),
\end{equation}	
where $\bbA_{\cdot i}$ denotes the $i$th column of the adjacency matrix $\bbA$, and $\bbA_{i\cdot}$ the $i$th row.

The algorithm starts from a random unit-norm vector and then via \eqref{e:feasible_update_rule} it takes a descent path towards a stationary point. The overall procedure is tabulated under Algorithm~\ref{alg:feasible_f_max}. It is often prudent to run the iterations multiple times using random initializations, and retain the solution that yields the least cost. Although Algorithm~\ref{alg:feasible_f_max} only guarantees convergence to a stationary point of the directed variation cost, in practice we have observed that it tends to find $f_{\max}=\text{DV}(\bbu_{\max})$ exactly if the number of initializations is chosen large enough; see Section \ref{S:numerical}. While finding $f_{\max}$ is of interest in its own right, our focus next is on using the obtained $\bbu_{\max}$ to formulate and solve the spectral dispersion minimization problem \eqref{eq:delta_opt_prob}.

\subsection{Spectral dispersion minimization}
As the second and final step, here we develop an algorithm to find the orthonormal basis $\bbU$ that minimizes the spectral dispersion \eqref{e:dispersion_def}. To cast the optimization problem \eqref{eq:delta_opt_prob} in the form of \eqref{e:general_feasible_orthogonal} and apply the previously outlined feasible method, we penalize the objective $\delta(\bbU)$ with a measure of the constraint violations to obtain
\begin{align}
\label{e:opt_dispersion_feasible}
& \min_{\bbU}
& & \hspace{-1cm} \phi(\bbU):=\delta(\bbU)+\frac{\lambda}{2}\left(\| \bbu_1-\bbu_{\min}\|^2+\| \bbu_N-\bbu_{\max}\|^2\right) \nonumber\\
& \text{subject to}
& &  \hspace{-0.2cm}\bbU^T\bbU=\bbI_N,
\end{align}
where $\lambda>0$ is chosen large enough to ensure $\bbu_1 = \bbu_{\min}$ and $\bbu_N = \bbu_{\max}$. \blue{Since we can explicitly monitor whether the constraints are enforced, the choice of $\lambda$ is often made via trial and error; see also Section~\ref{S:numerical} for an empirical demonstration.}
The resulting iterations are tabulated under Algorithm \ref{alg:feasible_dispersion}, where the gradient matrix $\bbG := \nabla\phi(\bbU) \in \reals^{N \times N}$ has columns given by
\begin{align}\label{e:grad_dispersion}
\bbg_1 & =  \left[\text{DV}(\bbu_{1})-\text{DV}(\bbu_{2})\right]\bar{\bbg}(\bbu_{1}) + \lambda(\bbu_{1} - \bbu_{\min}) \nonumber \\
\bbg_i & = \left[2\text{DV}(\bbu_{i})-\text{DV}(\bbu_{i+1})-\text{DV}(\bbu_{i-1})\right]\bar{\bbg}(\bbu_{i}),\: 1 < i < N \nonumber \\
\bbg_N & = \left[\text{DV}(\bbu_{N-1})-\text{DV}(\bbu_{N})\right]\bar{\bbg}(\bbu_{N}) + \lambda(\bbu_{N} - \bbu_{\max}),
\end{align}
where the entries of $\bar{\bbg}$ are specified in \eqref{e:f_max_gradient}. 
Once more, it is convenient to run the algorithm multiple times and retain the least disperse DGFT basis $\hbU$.

\begin{algorithm}[t]
	\caption{Spectral Dispersion Minimization}
	\label{alg:feasible_dispersion}
	\begin{algorithmic}[1]
		\STATE \textbf{Input:} Adjacency matrix $\bbA$, parameters $\lambda > 0$ and $\epsilon > 0$.
		\STATE Find $\bbu_{\max}$ using Algorithm \ref{alg:feasible_f_max} and set $\bbu_{\min}=\frac{1}{\sqrt{N}}\mathbf{1}_{N}$.
		\STATE \textbf{Initialize} $k=0$ and orthonormal $\bbU_{0} \in \reals^{N \times N}$ at random.
		\REPEAT
		\STATE Evaluate objective $\phi(\bbU_{k})$ in \eqref{e:opt_dispersion_feasible}.
		\STATE Compute gradient $\bbG_{k} \in \reals^{N \times N}$ via \eqref{e:grad_dispersion}.
		\STATE Form $\bbB_{k} = \bbG_{k} \bbU_{k}^{T} - \bbU_{k} \bbG_{k}^{T}$.
		\STATE Select $\tau_{k}$ satisfying conditions \eqref{e:Armijo-Wolfe_conditionsa} and \eqref{e:Armijo-Wolfe_conditionsb}.
		\STATE Update $\bbU_{k+1} (\tau_{k}) = (\bbI_N+\frac{\tau_{k}}{2} \bbB_{k})^{-1}(\bbI_N - \frac{\tau_{k}}{2} \bbB_{k})\bbU_{k}$.
		\STATE $k \gets k+1$.
		\UNTIL{$\lVert \bbU_{k}- \bbU_{k-1} \rVert_{F} \leq \epsilon$}
		\STATE \textbf{Return} $\hbU=\bbU_{k}$.
	\end{algorithmic}
\end{algorithm}
While provably convergent to a stationary point of \eqref{e:opt_dispersion_feasible}, Algorithm~\ref{alg:feasible_dispersion} does not offer guarantees on the global optimality of the solution $\hbU$. Still, numerical tests in Section~\ref{S:numerical} corroborate the effectiveness of the proposed optimization strategy \blue{and its robustness with respect to the initialization}. The computational complexity of Algorithm~\ref{alg:feasible_dispersion} is $\mathcal{O}(N^3)$ per iteration due to the matrix inversion involved in the calculation of the Cayley transform. In the next section we propose a lightweight heuristic to construct spread DGFT basis using the eigenvectors of 
$\ccalG^u$'s Laplacian matrix. \blue{But before moving on, a remark on the relationship between the DGFT and the DFT of discrete-time signals is in order.

\begin{remark}[Relationship with the DFT] \label{remark:DFT_rel}\normalfont
The scope of the proposed DGFT framework [in particular the notion of directed variation in \eqref{e:DV_def}] is limited to real-valued signals and basis vectors. Accordingly, for the directed cycle graph in Fig. \ref{fig:f_max_special}-(b) which represents the support of periodic discrete-time signals, one would fail to obtain the (complex-valued) DFT as the solution of \eqref{eq:delta_opt_prob}. Still, one can modify the definition of directed variation in \eqref{e:DV_def} to recover the classical DFT when $\ccalG$ is a directed cycle. To this end, for complex-valued $\bbx$ consider
	\begin{equation} \label{e:DV_def_DFT}
		\text{DV}_{\text{DFT}}(\bbx) \coloneqq \frac{1}{\sum_{j,k=1}^{N} A_{jk}} \sum_{\substack{j,k=1}}^{N} A_{jk}[\text{mod}_{2\pi}(\omega_j - \omega_k)],
	\end{equation}
where $i=\sqrt{-1}$ is the imaginary number and $\omega_j \in [0,2\pi)$ denotes the phase angle of $x_j$, the signal value at node $j$. Using this definition, one 
can formulate the following complex-valued spectral dispersion minimization problem [cf. \eqref{eq:delta_opt_prob}]
%
\begin{align}
\label{eq:delta_opt_prob_DFT}
& \min_{\bbU} 
& & \hspace{-0.2cm}\sum_{r=1}^{N} \left[	\text{DV}_{\text{DFT}}(\bbu_{r+1})-	\text{DV}_{\text{DFT}}(\bbu_r)\right]^2 \\
& \text{subject to}
& & \hspace{-0.1cm} \bbU^{H}\bbU=\bbI_N,\nonumber\\
&
&&\hspace{-0.1cm} \bbu_1=\frac{1}{\sqrt{N}}\mathbf{1}_{N},\nonumber\\
&
&& \hspace{-0.1cm}\text{DV}_{\text{DFT}}(\bbu_{N+1})=2 \pi,\nonumber
\end{align}
where $^{H}$ stands for conjugate transposition. Notice that $\bbu_{N+1}$ is not an optimization variable, and the third constraint fixes $[2\pi - \text{DV}_{\text{DFT}}(\bbu_N)]^2$ as the last (i.e., $r=N$) summand of the objective function. 

When $\ccalG$ is the unweighted, directed cycle with $N$ nodes, the global optimum $\bbU_{\text{DFT}}^{*}$ of \eqref{eq:delta_opt_prob_DFT}  is the $N$-point DFT matrix, with entries $U_{\text{DFT},jk}^{*}=\frac{1}{\sqrt{N}} e^{\frac{-2\pi i j k}{N}}$. The frequencies form an arithmetic sequence, while covering the whole frequency range $[0,2\pi)$; see Fig.~\ref{fig:DFT}. It is fair to say that the objective function of \eqref{eq:delta_opt_prob_DFT} does not subsume its counterpart \eqref{e:dispersion_def} as a special case for real signals. While an iterative solver is not needed in this particular case,  the feasible method in~\cite{wen2013feasible} can accommodate complex-valued Stiefeld manifold constraints.
\end{remark}

\begin{figure}[t]
	\centering    
	{\includegraphics[width=1\linewidth]{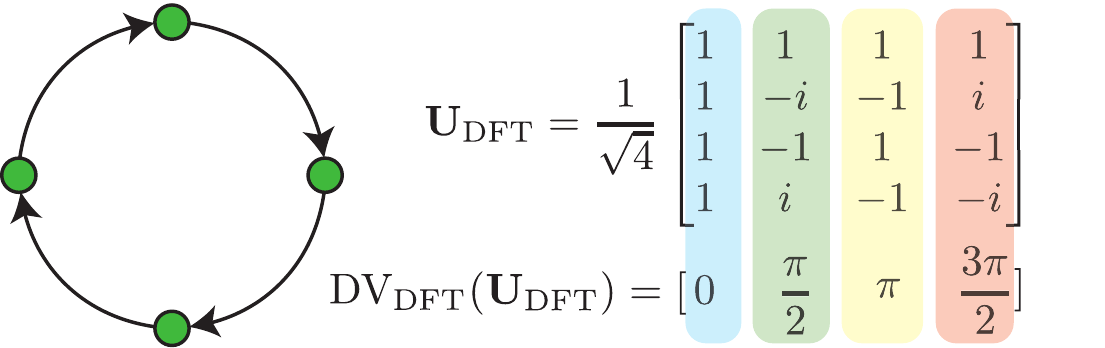}}
	\caption{ {(left) A directed cycle graph with $N=4$ nodes.  (right) The $4$-point DFT basis is recovered as the optimal solution of \eqref{eq:delta_opt_prob_DFT}}.}
	\label{fig:DFT}
\end{figure}
}
\section{A Digraph Fourier Transform Heuristic}\label{S:heuristic}
As an alternative to the feasible method discussed in the previous section, here we consider the underlying undirected graph $\ccalG^u$ and use the eigenvectors of its Laplacian matrix to construct a disperse set of frequencies. The reason for using the eigenvectors of $\bbL$ is that i) they are widely used for undirected graphs having good localization properties in the vertex domain; ii) we can modify the frequencies (in the digraph $\ccalG$) by flipping the sign of each Laplacian eigenvector; and iii) they can be used to approximate $f_{\max}$ within a factor of $1/2$ as asserted in Proposition \ref{pro:f_approx}. Unlike the undirected case, the directed variation of any eigenvector $\bbu$ will in general be different from the variation of $-\bbu$; so we can pick the one that we desire without compromising the orthogonality constraint. 

Fixing $f_1=0$ and $f_N=\tilde{f}_{\text{max}}$ from \eqref{eq:f_tilde}, in lieu of \eqref{e:directed_freq} we will henceforth construct a disperse set of frequencies by using the eigenvectors of $\bbL$. Let $f_i\coloneqq \text{DV}(\bbu_i)$ and $\overline{f}_i\coloneqq \text{DV}(-\bbu_i)$, where $\bbu_i$ is the $i$th eigenvector of $\bbL$. Define the set of all {candidate} frequencies {as} $\ccalF:=\{f_i,\overline{f}_i:1<i<N\}$. {The goal is to select} $N-2$ frequencies from $\ccalF$ such that together with $\{0,\tilde{f}_{\text{max}}\}$ they form our well-spread Fourier frequencies. To preserve orthonormality, we would select exactly one from each pair $\{f_i,\overline{f}_i\}$. We will argue later that this induces a matroid basis constraint on a supermodular (frequency) set minimization problem described next.

\subsection{Frequency selection via supermodular minimization}
 {To find the DGFT basis in accordance to the design criterion in Section \ref{S:s_problem}, we define a spectral dispersion (set) function} that measures how well spread the corresponding frequencies are over $[0,\tilde{f}_{\text{max}}]$. For a candidate {frequency set} $\ccalS\subseteq \ccalF$, let $s_1\leq s_2\leq ...\leq s_m$ be the elements of $\ccalS$ in non-decreasing order, where $m=\abs{\ccalS}$. Then we define the dispersion of $\ccalS$ as
\begin{equation} \label{e:dispersion_def_set}
\delta(\ccalS)=\sum_{i=0}^{m} (s_{i+1}-s_i)^2,
\end{equation}
where $s_0=0$ and $s_{m+1}=\tilde{f}_{\text{max}}$ [cf. \eqref{e:dispersion_def}].
It can be verified that $\delta(\ccalS)$ is a monotone non-increasing function, which means that for any sets $\ccalS_1\subseteq \ccalS_2$, we have $\delta(\ccalS_1)\geq \delta(\ccalS_2)$. For a fixed value of $m$, one can show that $\delta(\ccalS)$ is minimized when the $s_i$'s form an arithmetic sequence, {consistent} with our design goal in \eqref{e:directed_freq}. Hence, we {seek} to minimize $\delta(\ccalS)$ through a set function optimization procedure. In Lemma \ref{lemma:obj} we show that the dispersion function (\ref{e:dispersion_def_set}) has the supermodular property. First, for completeness we define submodularity/supermodularity.
\begin{mydefinition}[Submodularity]\normalfont
Let $\ccalS$ be a finite ground set. A set function $f:2^{\ccalS}\mapsto \reals$ is submodular if:
\begin{equation}
\label{eq:def_submod}
f(\ccalT_1\cup \{e\})-f(\ccalT_1)\geq f(\ccalT_2\cup \{e\})-f(\ccalT_2),
\end{equation}
for all subsets $\ccalT_1\subseteq \ccalT_2\subseteq \ccalS$ and any element $e\in \ccalS\backslash \ccalT_2$.
\end{mydefinition}
Equation \eqref{eq:def_submod} is also known as the \emph{diminishing returns} property. It means that adding a single element $e$ results in less gain when added to a bigger set $\ccalT_2$, compared to adding the same element to a subset of $\ccalT_2$ like $\ccalT_1$. The diminishing returns property arises in many science and engineering applications including facility location, sensor placement, and feature selection \cite{nemhauser1978analysis}, where adding a new sensor/feature or opening a new location becomes increasingly less beneficial as one has more and more of them already available. A set function $f$ is said to be supermodular if $-f$ is submodular, i.e. \eqref{eq:def_submod} holds in the other direction. Roughly speaking, for supermodular functions, items have more value when bundled together.
\begin{mylemma}
\label{lemma:obj}\normalfont
The {spectral} dispersion function $\delta:2^\ccalF\mapsto \reals$ defined in (\ref{e:dispersion_def_set}) is a supermodular function.
\end{mylemma}
\begin{myproof}
Consider two subsets $\ccalS_1,\ccalS_2$ such that $\ccalS_1\subseteq \ccalS_2\subseteq \ccalF$, and a single element $e\in \ccalF\backslash \ccalS_2$. Let $s_1^L$ and $s_1^R$ be the largest value smaller than $e$ and the smallest value greater than $e$ in $\ccalS_1\cup\{0,\tilde{f}_{\max}\}$, respectively {(i.e.,} $e$ breaks the gap between $s_1^L$ and $s_1^R$). Similarly, let $s_2^L$ and $s_2^R$ be defined for $\ccalS_2$. Since $\ccalS_1\subseteq \ccalS_2$, then $s_1^L\leq s_2^L\leq e\leq s_2^R\leq s_1^R$. {The result follows by comparing} the marginal values
	\begin{align*}
	\delta(\ccalS_1\cup \{e\})-\delta(\ccalS_1)={}&(s_1^R-e)^2+(e-s_1^L)^2-(s_1^R-s_1^L)^2\\
	={}&-2(s_1^R-e)(e-s_1^L)\\
	\leq{}& -2(s_2^R-e)(e-s_2^L)\\
={}&\delta(\ccalS_2\cup \{e\})-\delta(\ccalS_2).
	\end{align*}
\end{myproof}

{Recalling the orthonormality constraint, we define} $\ccalB$ to be the set of all subsets $\ccalS\subseteq \ccalF$ that satisfy $\abs{\ccalS\cap \{f_i,\overline{f_i}\}}=1$, $i=2,...,N-1$. Then, {frequency selection from $\ccalF$ boils down to solving}
\begin{equation}
 {\min_{\ccalS}  \delta(\ccalS), \quad 
 \text{subject to } \ccalS\in \ccalB.}
\label{eq:supmin}
\end{equation}
{Next}, in Lemma \ref{lemma:constraint} we show that the constraint in \eqref{eq:supmin} is a matroid basis constraint. {To state that result, we first define the notions of matroid and partition matroid.}
\begin{mydefinition}[Matroid]\normalfont
Let $\ccalS$ be a finite ground set and let $\ccalI$ be a collection of subsets of $\ccalS$. The pair $\ccalM=(\ccalS,\ccalI)$ is a matroid if the following properties hold:
\begin{itemize}
\item Hereditary Property: If $\ccalT\in \ccalI$, then $\ccalT'\in \mathcal{I}$ for all $\ccalT'\subseteq \ccalT$.
\item Augmentation Property: If $\ccalT_1,\ccalT_2\in \mathcal{I}$ and $\abs{\ccalT_1}<\abs{\ccalT_2}$, then there exists $e\in \ccalT_2\backslash \ccalT_1$ such that $\ccalT_1\cup \{e\}\in \mathcal{I}$.
\end{itemize}
The collection $\mathcal{I}$ is called the set of independent sets of the matroid $\mathcal{M}$. A maximal independent set is a basis. One can show that all the \blue{basis vectors} of a matroid have the same cardinality.
\end{mydefinition}
A matroid is a powerful structure used in combinatorial optimization, generalizing the notion of linear independence in vector spaces. Indeed, it is not hard to observe that if $\ccalS$ is a set of (not necessarily independent) vectors, then the linearly independent subsets of $\ccalS$ form a valid independent family that satisfies the above two properties. The uniform matroid, graphic matroid, and partition matroid are other  examples of matroids. The latter one will be useful in the sequel.
%
\begin{mydefinition}[Partition matroid \cite{schrijver}]\normalfont
\label{def:partition_matroid}
Let $\ccalS$ denote a finite set and let $\ccalS_1,...,\ccalS_m$ be a partition of $\ccalS$, i.e. a collection of disjoint sets such that $\ccalS_1\cup ... \cup \ccalS_m=\ccalS$. Let $d_1,...,d_m$ be a collection of non-negative integers. Define a set $\ccalI$ by $\ccalA\in \ccalI$ iff $\abs{\ccalA\cap S_i}\leq d_i$ for all $i=1,...,m$. Then, $\ccalM=(\ccalS,\ccalI)$ is called the partition matroid.
\end{mydefinition}
All elements are now in place to establish that the orthonormality constraint in \eqref{eq:supmin} is a partition matroid basis constraint.
%
\begin{mylemma}
\label{lemma:constraint}\normalfont
There exists a (partition) matroid $\mathcal{M}$ such that the set $\mathcal{B}$ in \eqref{eq:supmin} is the set of all basis vectors of $\mathcal{M}$.
\end{mylemma}
\begin{myproof}
{Recall Definition \ref{def:partition_matroid} and set $\ccalS:=\ccalF$, $\ccalS_i:=\{f_i,\overline{f_i}\}$ and $d_i:=1$ for all $i=2,...,N-1$}, to get a partition matroid $\ccalM=(\ccalF,\ccalI)$. The \blue{basis vectors} of $\ccalM$, which are defined as the maximal elements of $\ccalI$, are those subsets $\ccalA\subseteq \ccalF$ that satisfy $\abs{\ccalA\cap \{f_i,\overline{f_i}\}}=1$ for all $i=2,...,N-1$, which are the elements of $\ccalB$.
\end{myproof}

\subsection{Greedy algorithm for DGFT \blue{basis} selection}
\noindent{Lemmas} \ref{lemma:obj} and \ref{lemma:constraint} {assert} that \eqref{eq:supmin} is a supermodular minimization problem subject to a matroid basis constraint. Since supermodular minimization is NP-hard and hard to approximate to any factor \cite{kelner2007hardness,mittal2013fptas}, we create a submodular function $\tilde{\delta}(\ccalS)$ and use the algorithms for submodular maximization to find a set of disperse \blue{basis} ${\bbU}$. In particular, we define
\begin{equation}
\tilde{\delta}(\ccalS):=\tilde{f}_{\text{max}}^2 - \delta(\ccalS),
\end{equation}
{which} is a non-negative (increasing) submodular function, because $\delta(\emptyset)=\tilde{f}_{\text{max}}^2$ is an upper bound for $\delta(\ccalS)$. There are several results for maximizing submodular functions under matroid constraints for {both the} non-monotone~\cite{lee2009non} and monotone cases \cite{fisher1978analysis,calinescu2011maximizing}. We {adopt} the greedy algorithm of \cite{fisher1978analysis} due to its simplicity ({tabulated} under Algorithm \ref{alg:greedy}), which provides a $1/2$-approximation guarantee (Theorem \ref{theorem:approximation}).
\begin{algorithm}[t]
\caption{Greedy Spectral Dispersion Minimization}
\label{alg:greedy}
\begin{algorithmic}[1]
	\STATE \textbf{Input:} Set of possible frequencies $\ccalF$.
	\STATE \textbf{Initialize} $\ccalS=\emptyset$.
	\REPEAT
		\STATE $e \gets \argmax_{f\in \ccalF} \left\{\tilde{\delta}(\ccalS\cup\{f\})-\tilde{\delta}(\ccalS)\right\}$.
		\STATE $\ccalS \gets \ccalS\cup \{e\}$.
		\STATE Delete from $\ccalF$ the pair $\{f_i,\overline{f}_i\}$ that $e$ belongs to.
	\UNTIL{$F=\emptyset$}
\end{algorithmic}
\end{algorithm}

The algorithm starts with an empty set $\ccalS$. In each iteration, it finds the element $e$ that produces the biggest gain in terms of increasing $\tilde{\delta}(\ccalS)$. Then it deletes the pair that $e$ belongs to, because the other element in that pair cannot be chosen by virtue of the matroid constraint. The running time of the algorithm is $\mathcal{O}(N^2)$, in addition to the $\mathcal{O}(N^3)$ cost of computing the Laplacian eigenvectors. 

\begin{mytheorem}[\hspace{1sp}\cite{fisher1978analysis}] \label{theorem:approximation}
Let $\ccalS^*$ be the solution of problem (\ref{eq:supmin}) and $\ccalS^{\text{g}}$ be the output of the greedy Algorithm~\ref{alg:greedy}. Then,
\begin{equation*}
\tilde{\delta}(\ccalS^{\text{g}})\geq\frac{1}{2}\times\tilde{\delta}(\ccalS^*).
\end{equation*}
\end{mytheorem}
Notice that Theorem \ref{theorem:approximation} offers a worst-case guarantee, and Algorithm \ref{alg:greedy} is usually able to find near-optimal solutions in practice.

In summary, the greedy DGFT {basis construction} algorithm entails the following steps. First, {we form $\ccalG^u$ and find} the eigenvectors of the graph Laplacian $\bbL$. Second, the set $\ccalF$ is formed by calculating the directed variation for each eigenvector {$\bbu_i$} and its negative {$-\bbu_i$, $i=2,\ldots,N-1$.} Finally, the greedy Algorithm~\ref{alg:greedy} is run on the set $\ccalF$, and the output determines the set of frequencies as well as the orthonormal set of DGFT \blue{basis vectors} comprising $\bbU$.

 \blue{
\begin{remark}[Computational complexity] \label{remark:comp_cost}\normalfont
While the heuristic method is computationally more efficient than the feasible method in Algorithm \ref{alg:feasible_dispersion}, it still requires a full diagonalization of the graph Laplacian matrix which costs $\mathcal{O}(N^3)$. The complexity of $\mathcal{O}(N^3)$ appears to be a bottleneck for large graphs in all the state-of-the-art existing methods \cite{sardellitti,DSP_freq_analysis,singh2016graph,girault2018gft}.  For particular cases speedups may be obtained by exploiting eigenvector routines for very sparse matrices, by relying on truncated decompositions (which could suffice to approximately decompose smooth signals), or e.g., through greedy approximate diagonalization to compute the Laplacian eigenbasis at lower cost~\cite{le2018approximate}. While certainly a very interesting and fundamental problem, there has been little progress to date when it comes to realizing the vision of a ``fast'' GFT, even for undirected graphs.
\end{remark}}
\section{Numerical Results}\label{S:numerical}
\begin{figure}[t]
	\centering    
	{\includegraphics[width=0.6\linewidth]{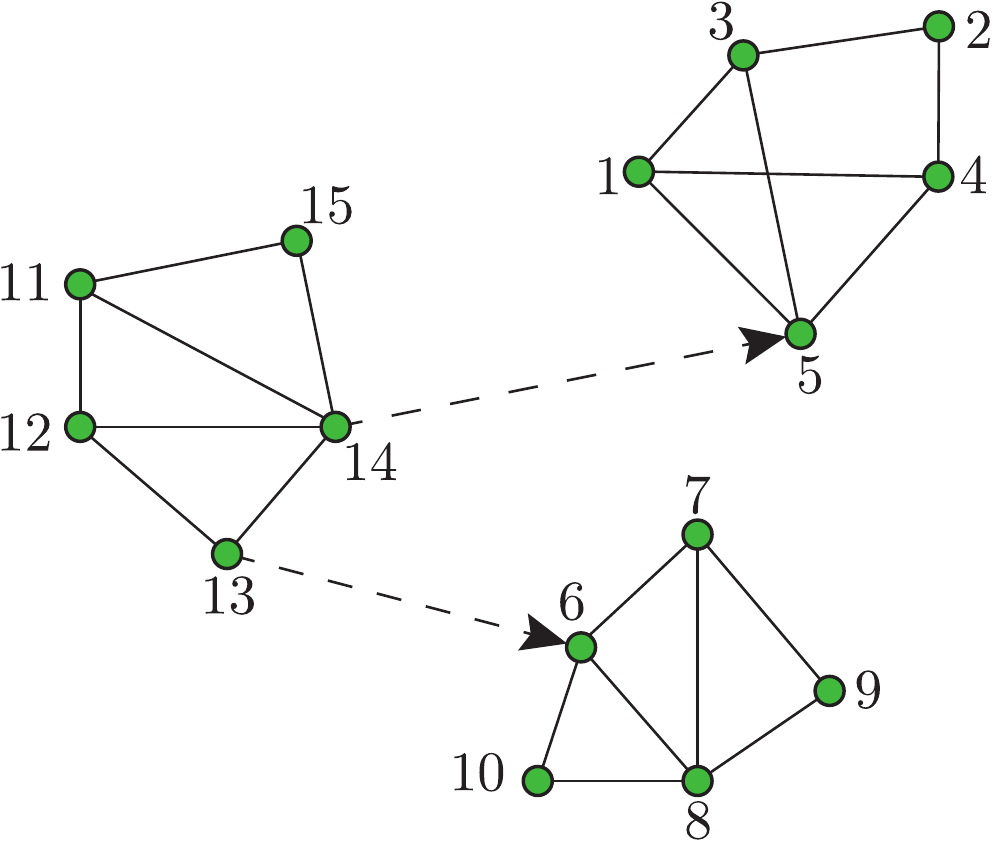}}
	\caption{Synthetic digraph with $N=15$ nodes and $2$ directed edges depicted as dashed arrows~\cite{sardellitti}.}
	\label{fig:graph_sardellitti}
\end{figure}

Here we carry out computer simulations on three graphs to assess the performance of the algorithms developed to construct a DGFT with spread frequency components. We also compare these basis signals with other state-of-the-art GFT methods.

\begin{figure*}[t]
	\centering    
	{\includegraphics[width=0.95\linewidth]{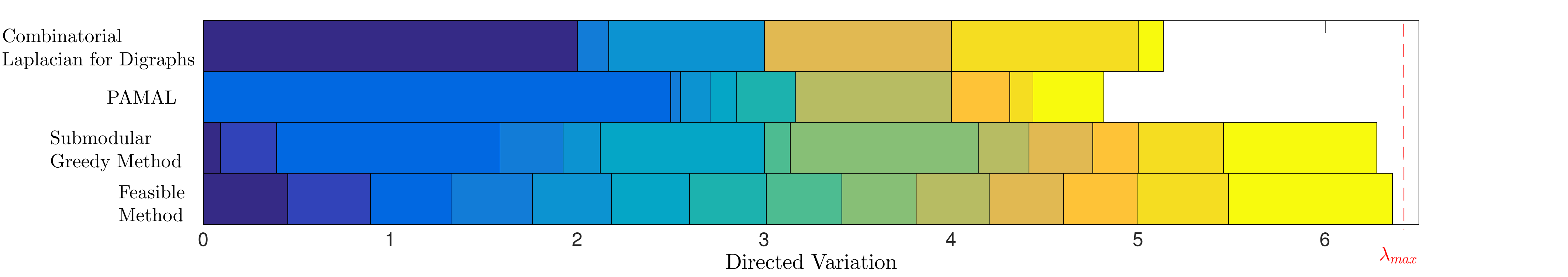}}
	\caption{Comparison of directed variations (i.e., graph frequencies) for the synthetic digraph in Fig. \ref{fig:graph_sardellitti} and different GFT methods: eigenvectors of the combinatorial Laplacian matrix $\bbL_d$ introduced in \cite{chung2005laplacians}; augmented Lagrangian method (PAMAL) in~\cite{sardellitti}; proposed greedy heuristic (Algorithm~\ref{alg:greedy}); and feasible method (Algorithm~\ref{alg:feasible_dispersion}). Colored boxes show the difference between two consecutive frequencies for each method, while the specific directed variation values correspond to the vertical boundary lines. Ideally one would like to have $N-1$ equal-sized boxes, but we argued that this is not always achievable. Notice how Algorithm~\ref{alg:feasible_dispersion} comes remarkably close to such a specification.}
	\label{fig:barplot}
\end{figure*}

\noindent\textbf{Synthetic digraph.} Using Algorithms \ref{alg:feasible_dispersion} and \ref{alg:greedy} we construct respective DGFTs for an unweighted digraph $\ccalG$ with $N = 15$ nodes shown in Fig.~\ref{fig:graph_sardellitti}, and compare them with the GFT put forth in \cite{sardellitti} that relies on an augmented Lagrangian optimization method termed PAMAL, as well as with the eigenbasis of the \blue{combinatorial Laplacian for directed graphs} in \cite{chung2005laplacians}. To define said \blue{combinatorial Laplacian for digraphs} $\bbL_d$, consider a random walk on the graph with transition probability matrix $\bbP=\bbD_{\text{out}}^{-1}\bbA$, where $\bbD_{\text{out}}$ is the diagonal matrix of node out-degrees. Let $\mathbf{\Pi}=\text{diag}(\bbpi)$ be the diagonal matrix with the stationary distribution $\bbpi$ of the random walk on the diagonal. Using these definitions, the \blue{combinatorial Laplacian for directed graphs} in~\cite{chung2005laplacians} is given by
%
$\bbL_d := \mathbf{\Pi}-(\mathbf{\Pi} \bbP + \bbP^T \mathbf{\Pi})/{2}.$
%

One would expect that the proposed DGFT approaches -- which directly optimize the spectral dispersion metric -- yield: i) a more spread set of graph frequencies; also ii) spanning a wider range of directed variations. This is indeed apparent from Fig. \ref{fig:barplot}, which depicts the distribution of frequencies (shown as vertical lines) for all GFT methods being compared. In particular, notice how the DGFT basis obtained via direct minimization of the dispersion cost (Algorithm~\ref{alg:feasible_dispersion}) yields an almost equidistributed set of graph frequencies. To further quantify this assertion, we first rescale the directed variation values to the $[0,1]$ interval and calculate their dispersion using \eqref{e:dispersion_def}. The results are reported in Table~\ref{tab:1}, which confirms that Algorithms~\ref{alg:feasible_dispersion} and \ref{alg:greedy} yield a better frequency spread (i.e., a smaller dispersion). While computationally more demanding, Algorithm~\ref{alg:feasible_dispersion} yields a more spread set of graph frequencies when compared to the greedy Algorithm~\ref{alg:greedy}, since it minimizes dispersion over a larger set [cf. \eqref{e:opt_dispersion_feasible} and \eqref{eq:supmin}]. Finally, Fig.~\ref{fig:feasible_subplots} shows the frequency components obtained via Algorithm~\ref{alg:feasible_dispersion}. Each subplot depicts one basis vector (column) of the resulting DGFT matrix $\bbU$, along with its corresponding directed variation values defined in \eqref{e:DV_def}. It is apparent that the first \blue{vectors} exhibit less variability than the higher frequency components. 
\blue{Moreover one can see that lower frequency components have the additional desired property of being roughly constant over network clusters; see also the design in~\cite{sardellitti}.}

\begin{figure}[t]
	\centering    
	{\includegraphics[width=0.95\linewidth]{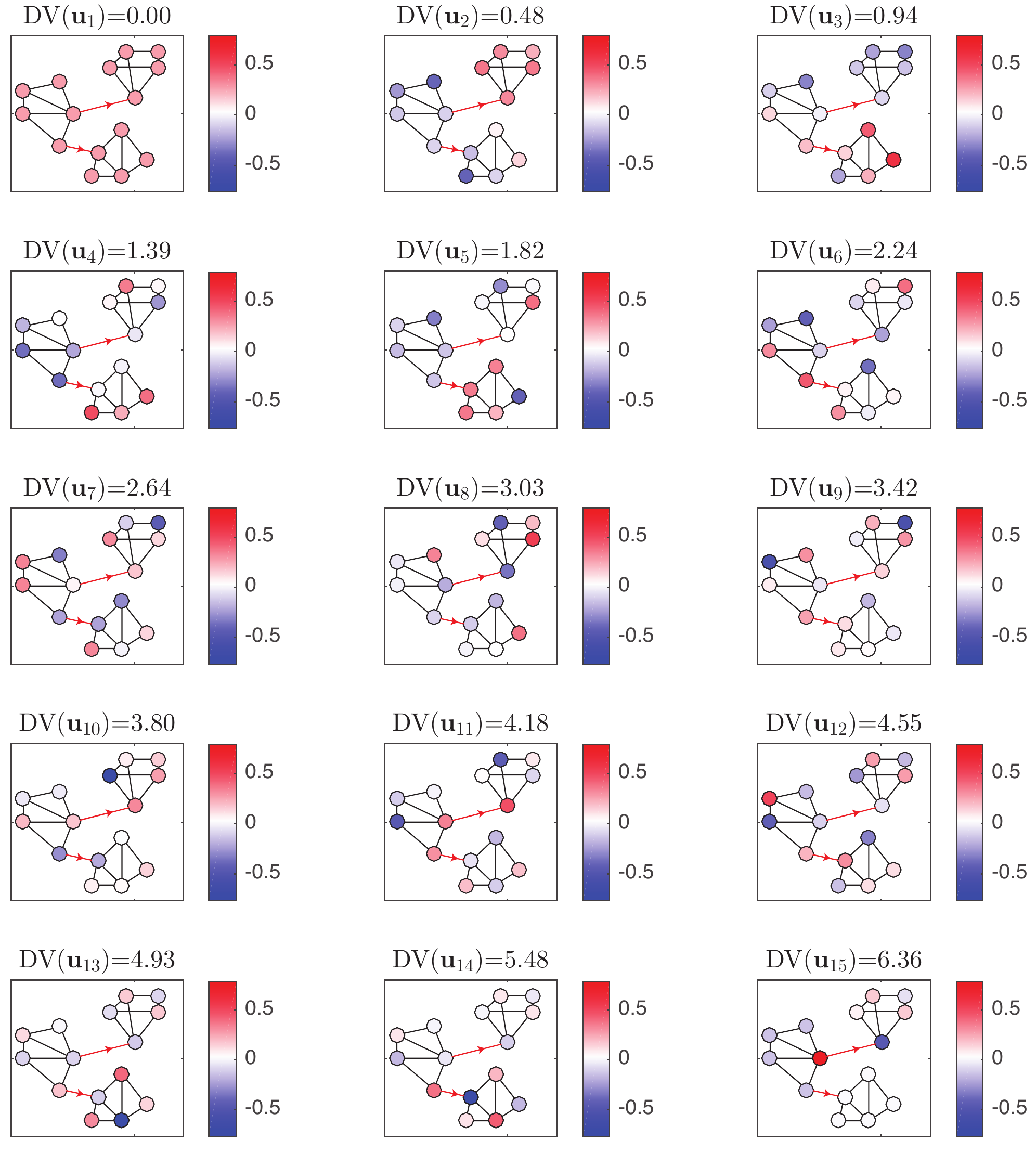}}
	\caption{DGFT \blue{basis vectors} obtained using Algorithm \ref{alg:feasible_dispersion} for the synthetic digraph in Fig. \ref{fig:graph_sardellitti}, along with their respective directed variation values (frequencies). Notice how the frequency components associated with lower frequencies are roughly constant over node clusters.}
	\label{fig:feasible_subplots}
	\vspace{-0.4cm}
\end{figure}

\begin{table}[t]
	\centering
	\footnotesize
	\begin{tabular}{|c|c|}
		
		\hline
		 \textbf{Method}	& \textbf{Dispersion}\\ \hline  \blue{Combinatorial Laplacian for directed graphs} \cite{chung2005laplacians} & 0.256 \\ \hline		PAMAL  \cite{sardellitti}  	& 0.301 \\ \hline	Submodular Greedy (Alg. \ref{alg:greedy}) 	&{\bf 0.118} \\  \hline Feasible Method (Alg. \ref{alg:feasible_dispersion}) &  {\bf 0.077} \\
		\hline
	\end{tabular}
	\caption{Spectral dispersion $\delta(\bbU)$ of obtained \blue{basis} $\bbU$ using different algorithms for the synthetic digraph in Fig. \ref{fig:graph_sardellitti}.}
	\label{tab:1}
	\vspace{-0.6cm}
\end{table}

\begin{figure}[t]
	\centering    
	{\includegraphics[width=\linewidth]{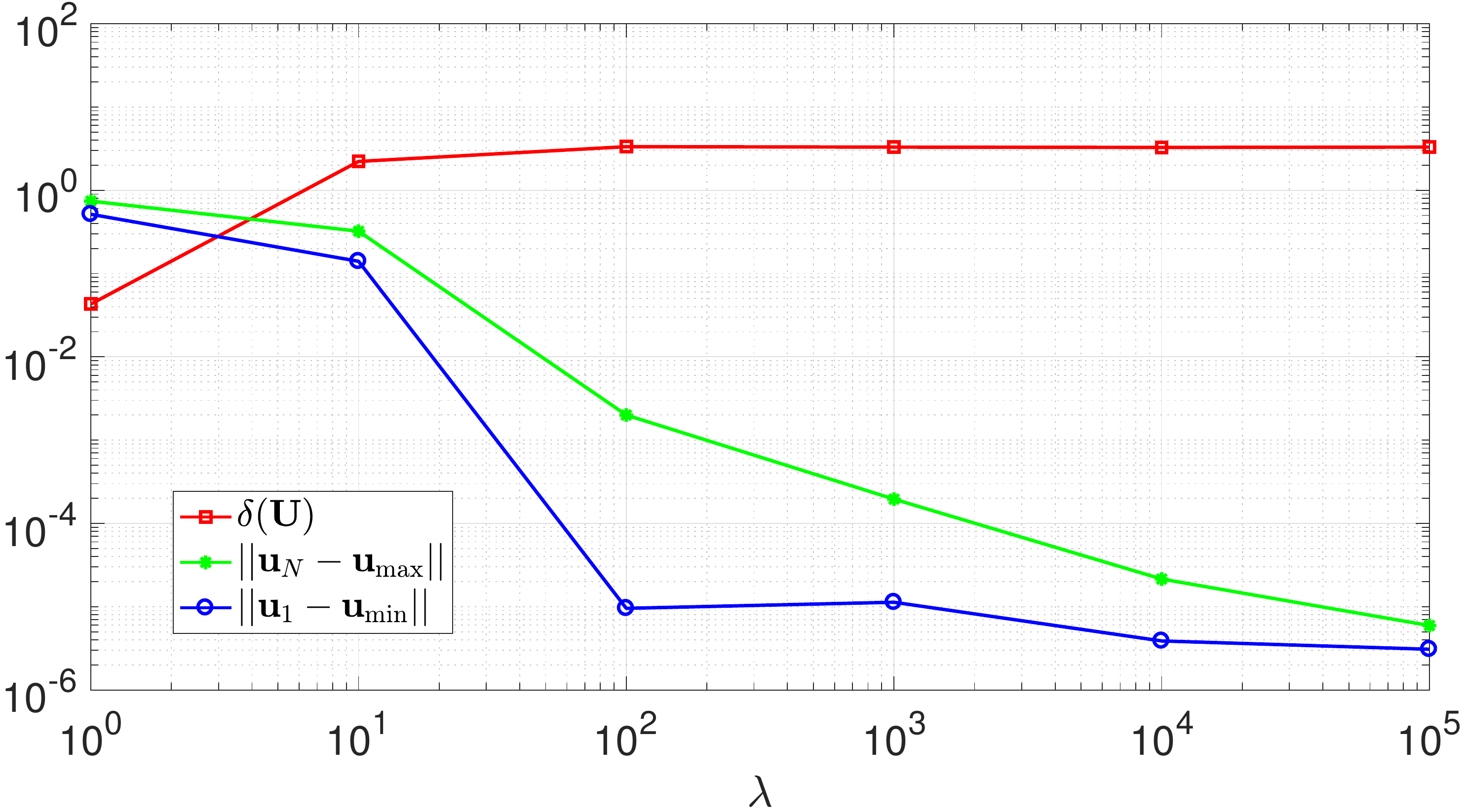}}
	\caption{\blue{Effect of $\lambda$ on the constraint violations $\|\bbu_1-\bbu_{\min}\|$, $\|\bbu_N-\bbu_{\max}\|$ and spectral dispersion $\delta(\bbU)$ in \eqref{e:opt_dispersion_feasible}.  We run Algorithm \ref{alg:feasible_dispersion} for the graph in Fig. \ref{fig:graph_sardellitti}, and average the results over $100$ Monte-Carlo simulations. For larger enough values of $\lambda$ (here $10^2$), we observe that the dispersion does not change and the last two constraints in \eqref{eq:delta_opt_prob} are (almost) satisfied.}}
	\label{fig:lambda_dependency}
\end{figure}
	
\begin{figure}[t]
	\begin{minipage}[b]{\linewidth}
		\centering
		\includegraphics[width=\textwidth]{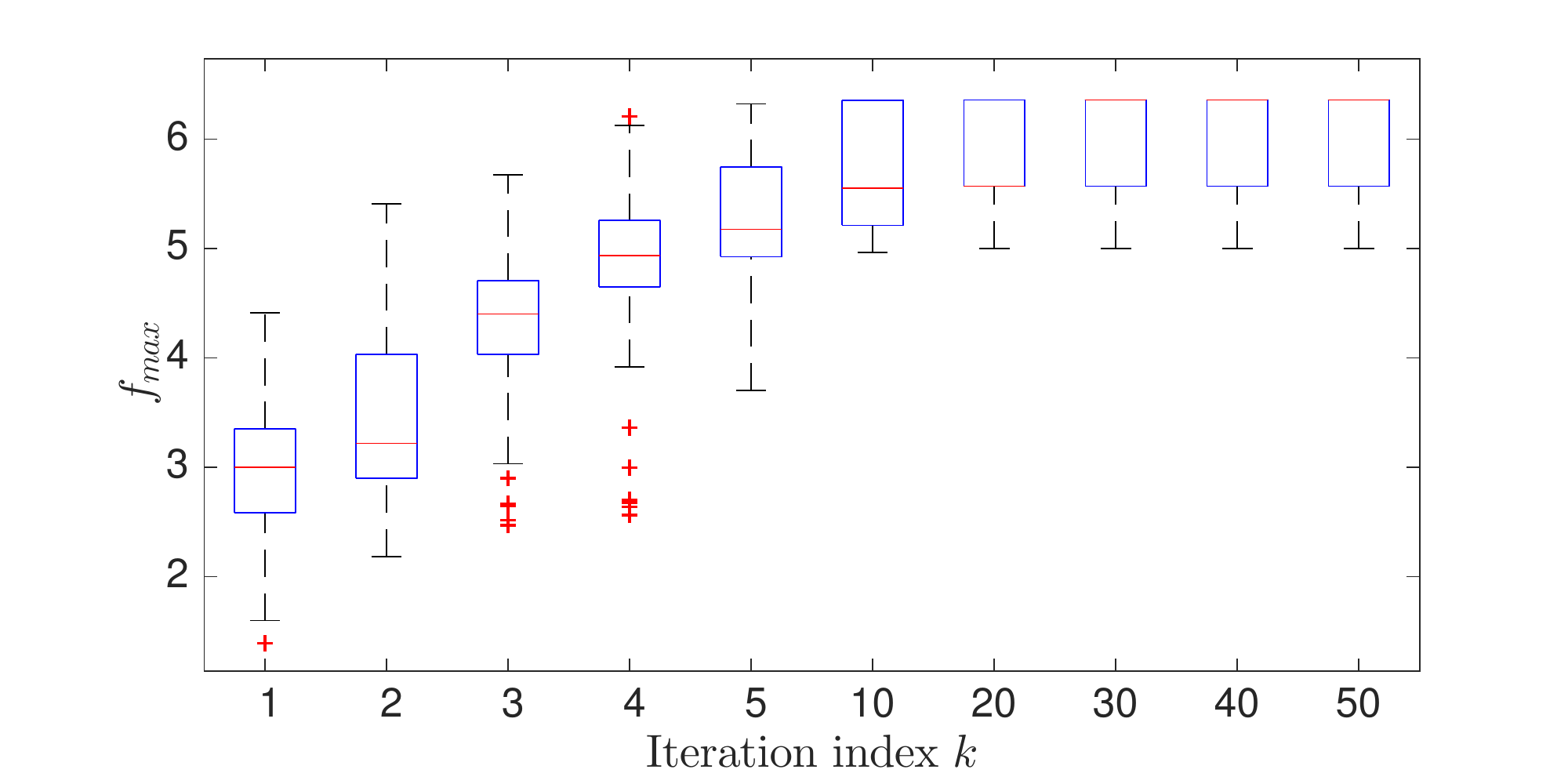}
	\end{minipage}
	\begin{minipage}[b]{\linewidth}
		\centering
		\includegraphics[width=\textwidth]{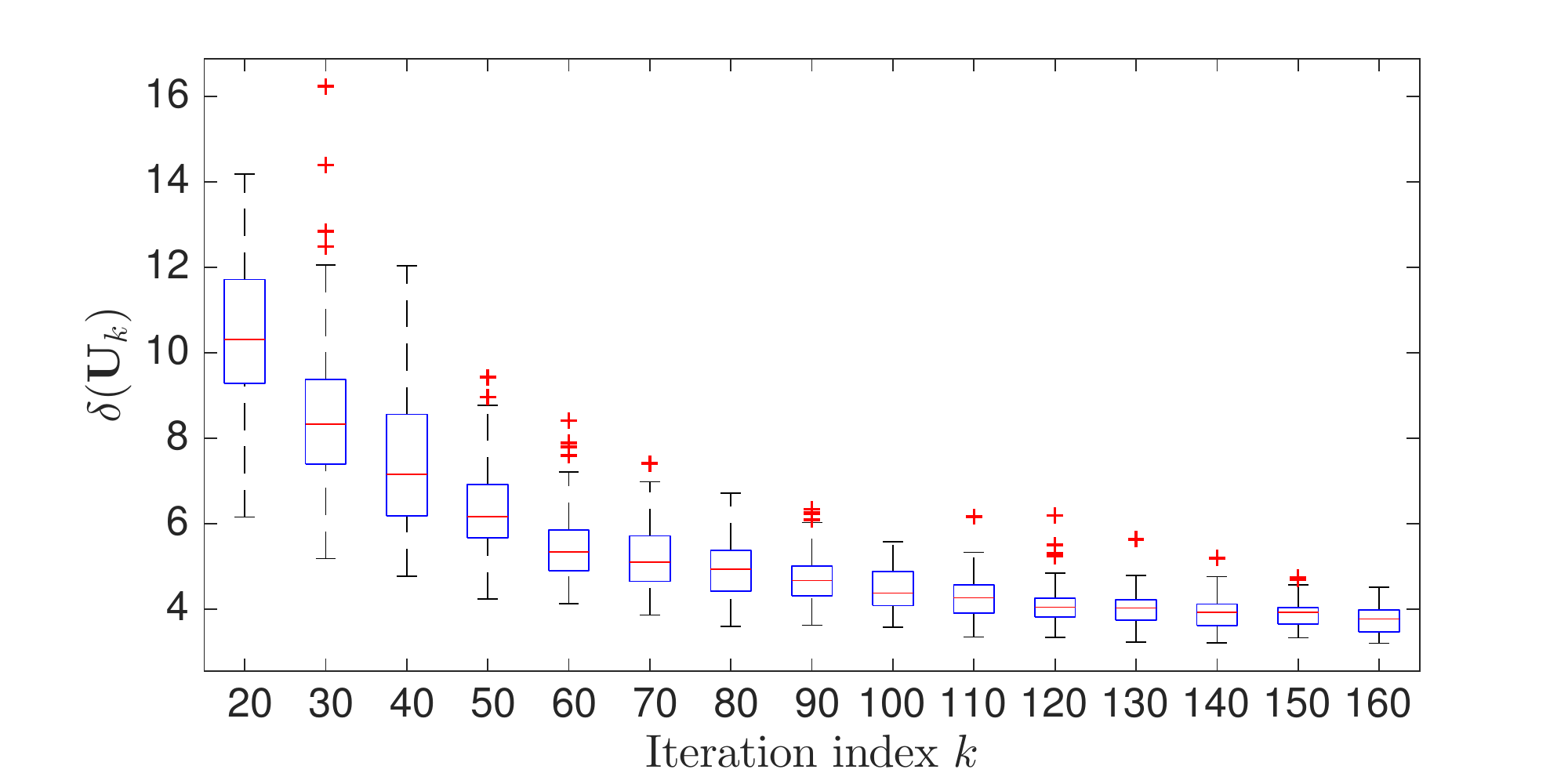}
	\end{minipage}
	\vspace{-0.5cm}	
	\caption{(top) Convergence behavior of Algorithm \ref{alg:feasible_f_max} for finding the maximum directed variation $f_{\max}$ of unit-norm signals on the synthetic digraph in Fig. \ref{fig:graph_sardellitti}. The boxes show the median and the 25th and 75th percentiles of $f_{\max}$ vs. the number of iterations $k$, obtained by running $100$ Monte-Carlo simulations based on independent initializations. (bottom) Likewise, but when using Algorithm \ref{alg:feasible_dispersion} to minimize the spectral dispersion $\delta(\bbU)$ in \eqref{e:dispersion_def}.}
	\label{fig:itr}
	\vspace{-0.3cm}
\end{figure}

We also use Monte-Carlo simulations to study \blue{the effect of $\lambda$ [cf. \eqref{e:opt_dispersion_feasible}]} on the convergence properties of our algorithms. \blue{After fixing $\bbu_{\max}$ and $\bbu_{\min}$, we run Algorithm~\ref{alg:feasible_dispersion} for different values of $\lambda$ to find the DGFT basis of the graph in Fig.~\ref{fig:graph_sardellitti}. The spectral dispersion in \eqref{e:dispersion_def} and the Euclidean distance between $\bbu_N(\bbu_1)$ and $\bbu_{\max} (\bbu_{\min})$ averaged over $100$ Monte-Carlo simulations are shown in Fig.~\ref{fig:lambda_dependency}. Apparently, for the extreme value of $\lambda = 0$ frequencies tend to be as close as possible, resulting in the smallest value of $\delta(\bbU)$. However, this solution is not feasible for problem \eqref{eq:delta_opt_prob}. By increasing $\lambda$ we trade-off dispersion for feasibility, making $\bbu_1$ and $\bbu_N$ closer to $\bbu_{\min}$ and $\bbu_{\max}$, respectively. This way, all constraints are satisfied resulting in a broader spectrum approximately spanning $[0,f_{\max}]$. For $\lambda>10^2$ in Fig~\ref{fig:lambda_dependency}, $\bbu_1$ and $\bbu_N$ become fixed and the remaining basis vectors spread as evenly as possible in the viable spectral band. Further exploring these trade-offs (including the localization properties of the resulting basis vectors) is certainly an interesting direction, which is beyond the scope of this paper and its DGFT design goals.} 

In Fig.~\ref{fig:itr} (top) we show the evolution of iterates for the feasible method in \cite{wen2013feasible}, when used to find the maximum directed variation (i.e., $f_{\max}$) for the same $15$-node graph in Fig. \ref{fig:graph_sardellitti}. We do so for $100$ different (random) initializations and report the median as well as the first and third quartiles versus the number of iterations. We observe that all the realizations converge~\cite[Theorem 2]{wen2013feasible}, but there is a small variation among the limiting values. This is expected because the feasible method is not guaranteed to converge to the global optimum of the non-convex problem \eqref{eq:delta_opt_prob}. It is worth mentioning that after about $10$ iterations, the exact value of $f_{\max}$ is achieved by a quarter of the realizations (and this improves to half of the realizations with about $30$ iterations).  Similarly, Fig.~\ref{fig:itr} (bottom) shows the median, first, and third quartiles of the dispersion function iterates $\delta(\bbU_k)$, when minimized using Algorithm~\ref{alg:feasible_dispersion}. Again, $100$ different Monte-Carlo simulations are considered and we observe that all of them \blue{robustly} converge to limiting values with small variability. This suggests that in practice we can run Algorithm~\ref{alg:feasible_dispersion} with different random initializations and retain the most spread frequency components among the obtained candidate solutions. 
%

\begin{figure*}[t]
	\includegraphics[width=1\linewidth]{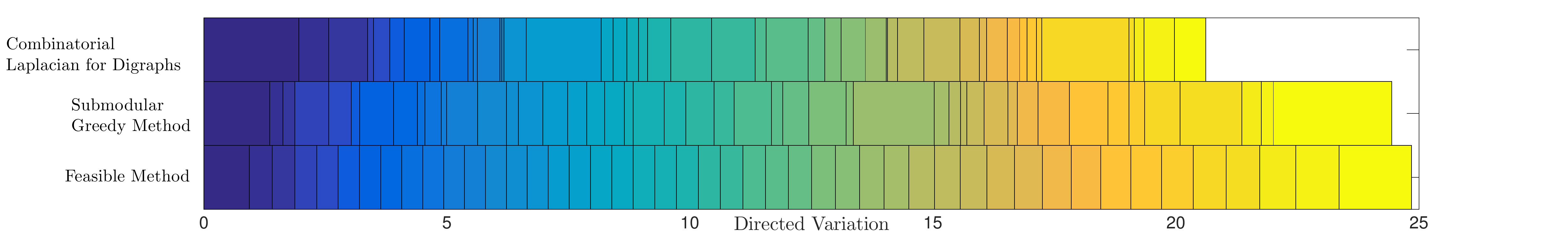}
	\caption{Comparison of directed variations (i.e., graph frequencies) for the structural brain network and different GFT methods: eigenvectors of the combinatorial Laplacian matrix $\bbL_d$ introduced in \cite{chung2005laplacians}; proposed greedy heuristic (Algorithm~\ref{alg:greedy}); and feasible method (Algorithm~\ref{alg:feasible_dispersion}). Colored boxes show the difference between two consecutive frequencies for each method, while the specific directed variation values correspond to the vertical boundary lines.}
	\label{fig:real_bar}
\end{figure*}

\begin{figure}[t]
	\centering    
	{\includegraphics[width=\linewidth]{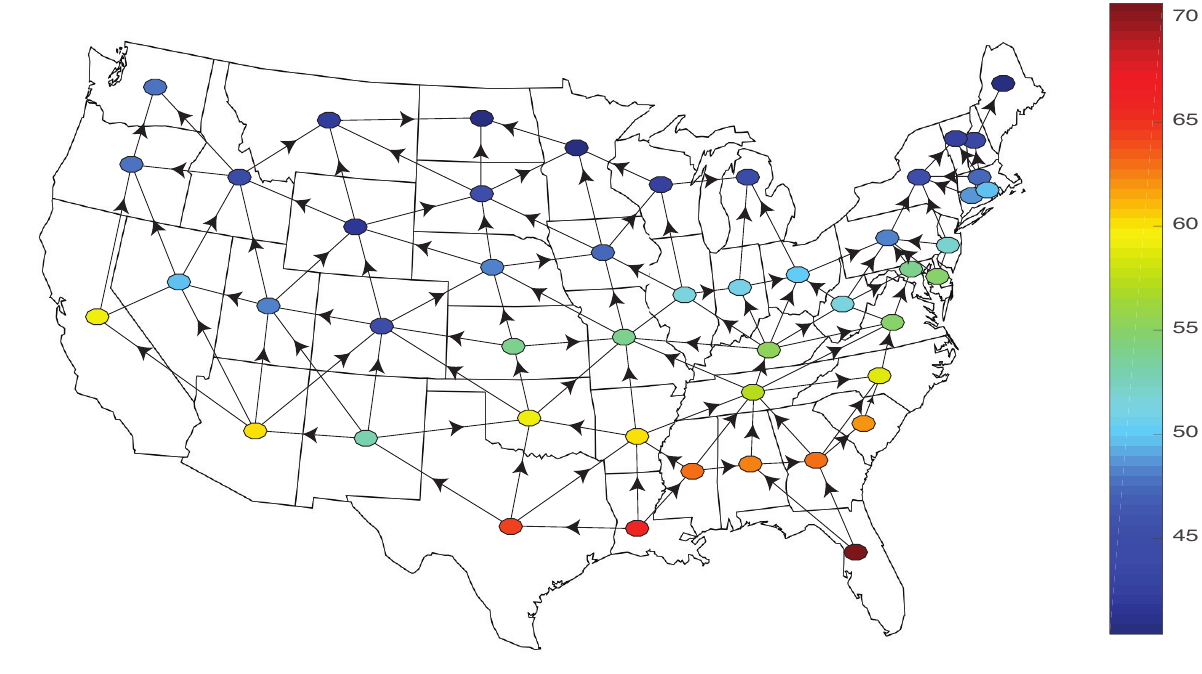}}
	\caption{Graph signal of average annual temperature in Fahrenheit for the contiguous US states. In the depicted digraph, a directed edge joins two states if they share a border, and the direction of the arc is set so that the state whose barycenter has lower latitude points to the one with higher latitude.}
	\label{fig:temp_signal}
	\vspace{-0.4cm}
\end{figure}
	
\noindent\textbf{Structural brain graph.} Next we consider a real brain graph representing the anatomical connections of the macaque cortex, which was studied in \cite{honey2007network,rubinov2010complex} for example. The network consists of $N=47$ nodes and 505 edges (among which 121 of them are directed).
The vertices represent different hubs in the brain, and the edges capture directed information flow among them. To corroborate that our resulting \blue{basis vectors} are well distributed in the graph spectral domain, Fig.~\ref{fig:real_bar} depicts the distribution of all the frequencies for the examined algorithms except for the PAMAL algorithm which did not converge within a reasonable time. In Fig.~\ref{fig:real_bar}, each vertical line indicates the directed variation (frequency) associated with a basis vector. Once more, the proposed algorithms are effective in terms of finding well dispersed and non-repetitive frequencies, which in this context could offer innovative alternatives for filtering of brain signals leading to potentially more interpretable graph frequency analyses~\cite{huang2016graph}. While certainly interesting, such a study is beyond the scope of this paper.

\begin{figure*}[t]
	\begin{minipage}[b]{0.33\textwidth}
		\centering
		\includegraphics[width=1\linewidth]{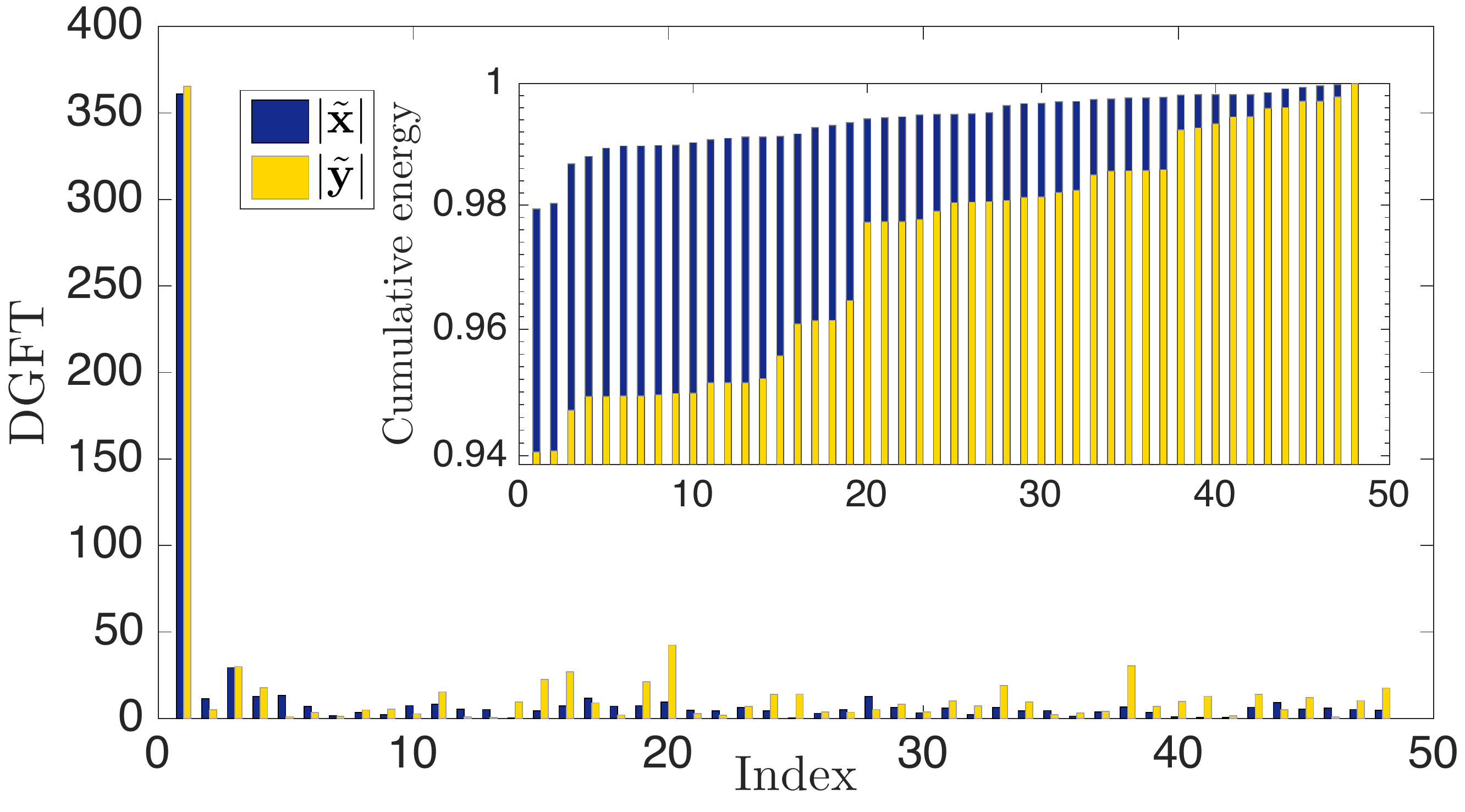}
		\centerline{(a)}\medskip
	\end{minipage}
	\hfill
	\begin{minipage}[b]{0.325\textwidth} 
		\centering
		\includegraphics[width=1\linewidth]{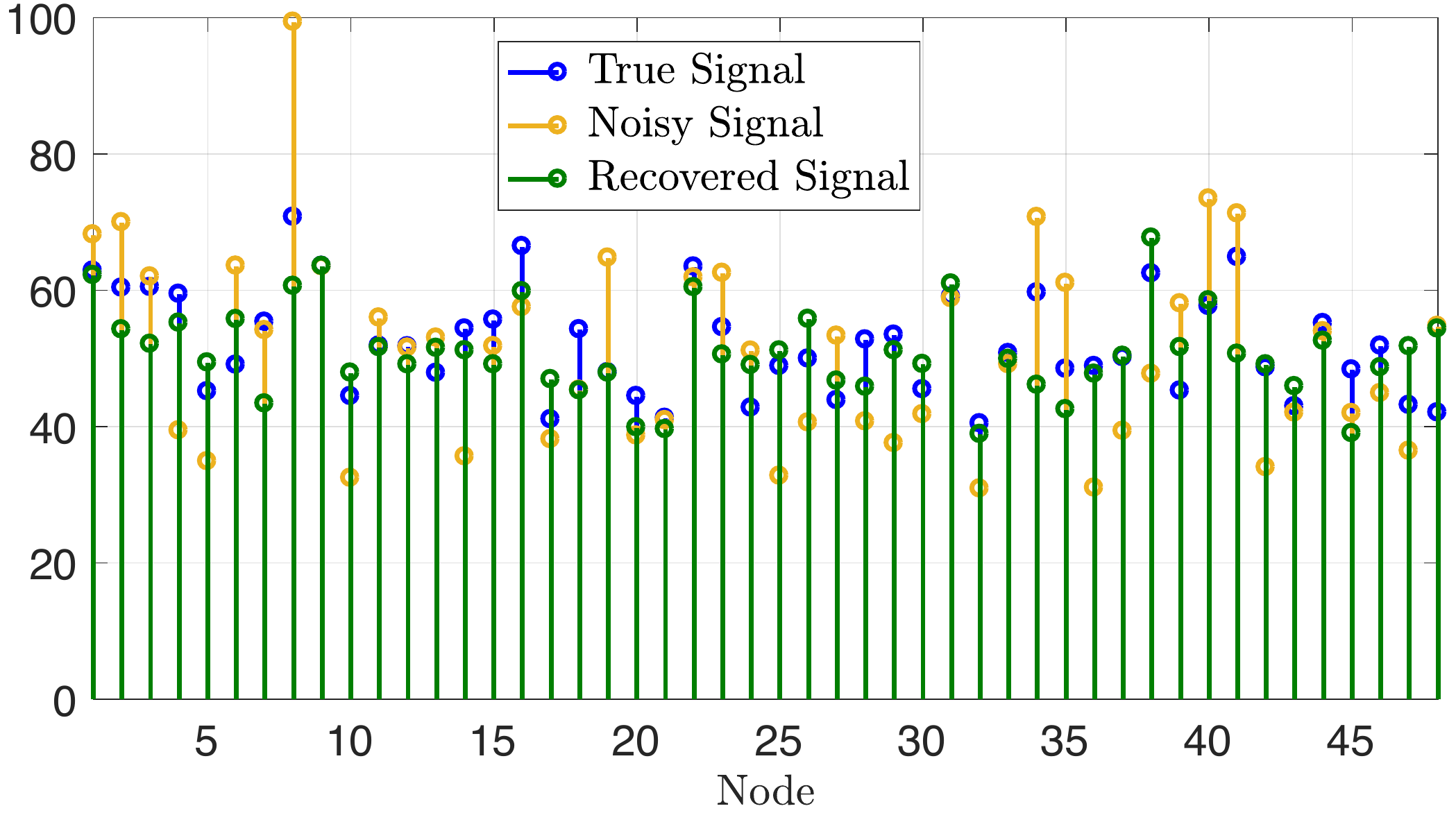}
		\centerline{(b)}\medskip
	\end{minipage}
	\hfill
	\begin{minipage}[b]{0.29\textwidth} 
		\centering
		\includegraphics[width=1\linewidth]{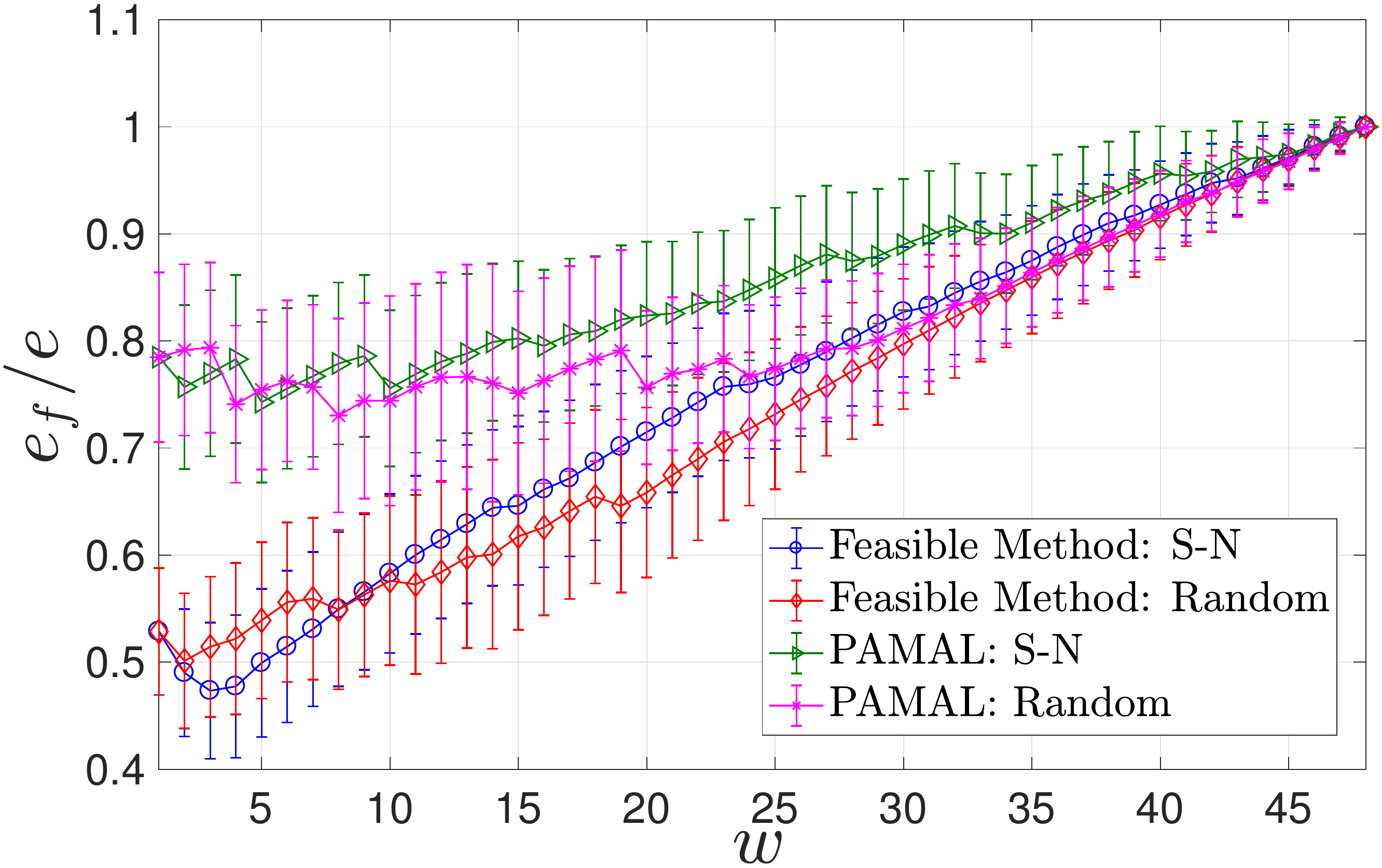}
		\centerline{(c)}\medskip
	\end{minipage}
	\vspace{-0.4cm}
	\caption{Denoising a temperature signal supported in the contiguous US digraph in Fig. \ref{fig:temp_signal}. (a)~DGFT of the original signal ($\tbx$) and the noisy signal ($\tby$), \blue{along with their cumulative energy distribution across frequencies.} (b)~A sample realization of the true, noisy, and recovered temperature signal \blue{for $w=3$}. (c)~ \blue{Ratio of} average recovery error using a low-pass filter to recovery error without filtering versus the window size  \blue{for the two digraphs with South to North (S-N) and random directed edges using the feasible method and PAMAL algorithm in \cite{sardellitti}.} As expected, \blue{on average the proposed method outperforms the PAMAL algorithm. Moreover, the lowest} error is obtained when filtering in the S-N digraph, which better captures the temperature flow from states with lower to higher latitudes.}
		\label{fig:filtering_feasible}
	\vspace{-0.3cm}
\end{figure*}

\begin{figure}[t]
	\centering    
	{\includegraphics[width=\linewidth]{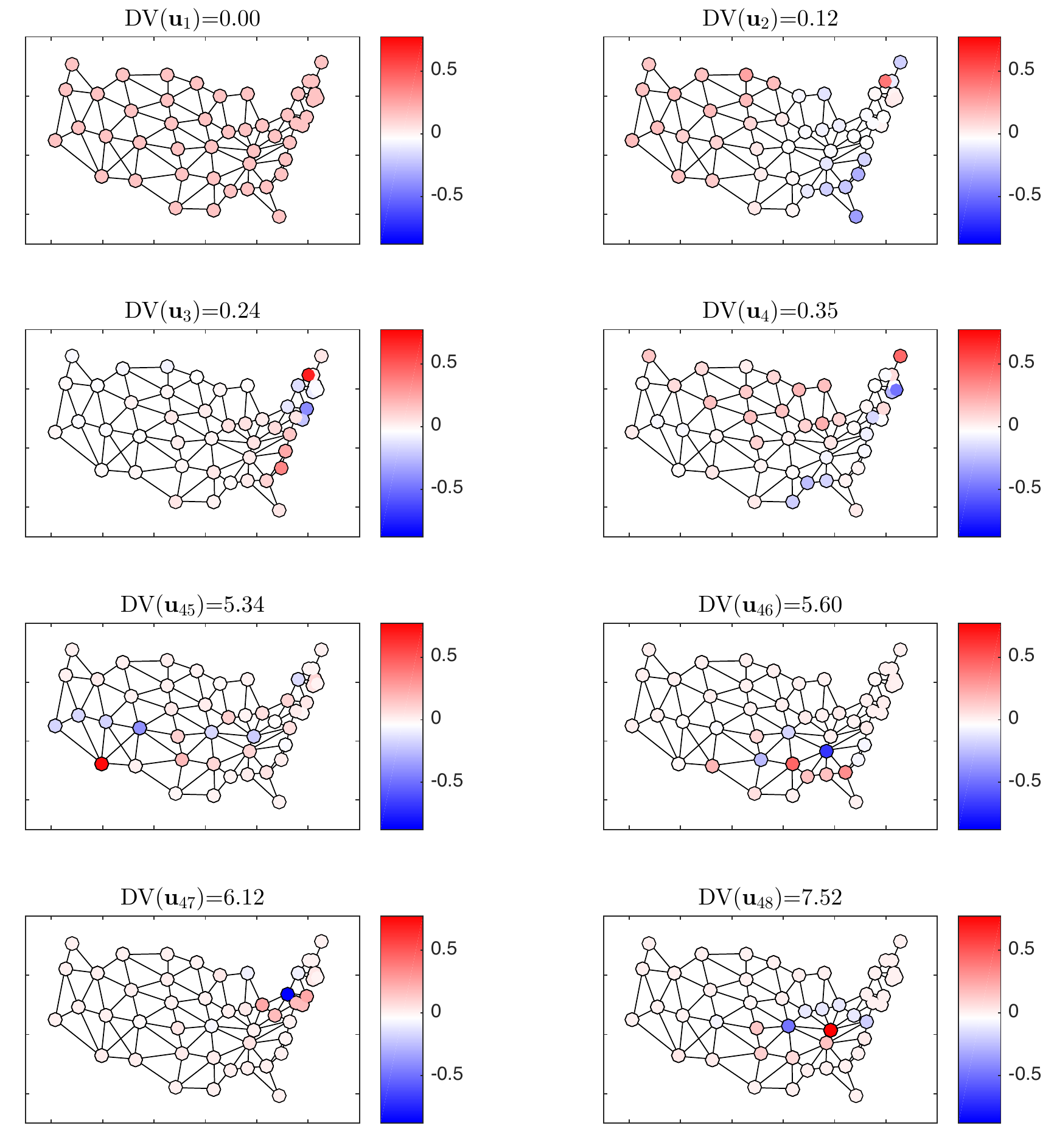}}
	\caption{DGFT \blue{basis vectors} obtained using Algorithm \ref{alg:feasible_dispersion} for the contiguous United States digraph in Fig. \ref{fig:temp_signal}, along with their respective directed variation values (frequencies). Eight frequency components are shown, corresponding to the lowest and highest four frequencies in the graph.}
	\label{fig:feasible_bases}
	\vspace{-0.6cm}
\end{figure}

\noindent\textbf{Contiguous United States.} Finally, we consider a digraph of the $N=48$ so-termed contiguous United States (excluding Alaska and Hawaii, which are not connected by land with the other states). A directed edge joins two states if they share a border, and the direction of the arc is set so that the state whose barycenter has a lower latitude points to the one with higher latitude, i.e., from South to North (S-N).  We also consider the average annual temperature of each state as the signal $\bbx\in\reals^{48}$ shown in Fig.~\ref{fig:temp_signal}.\footnote{Temperature data obtained from https://www.ncdc.noaa.gov} It is apparent from the temperature map that the states closer to the Equator (i.e., with lower latitude) have higher average temperatures. This justifies the adopted latitude-based graph construction scheme, to better capture a notion of flow through the temperature field. 

We determine a DGFT basis for this digraph via spectral dispersion minimization using Algorithm~\ref{alg:feasible_dispersion}. The resulting first and last $4$ frequency modes are depicted in Fig.~\ref{fig:feasible_bases}. The first four \blue{vectors} are smooth as expected. The last \blue{basis vectors} are smooth as well in the majority of the graph, but there exist a few nodes in them such that a highly connected vertex is significantly warmer than its northern neighbors, or, colder than its southern neighbors. For instance, in $\bbu_{48}$ Kentucky has markedly high temperature. This high-frequency modes can indeed help towards filtering out noisy measurements, as these spikes can be due to anomalous events or defective sensors. 

To corroborate this assertion, we aim to recover the temperature signal from noisy measurements $\bby = \bbx {+} \bbn$, where the additive noise $\bbn$ is a zero-mean, Gaussian random vector with covariance matrix $10 \bbI_{N}$. To that end, we use a low-pass graph filter with frequency response $\tilde{\bbh}=[\tilde{h}_{1},\ldots,\tilde{h}_{N}]^T$, where $\tilde{h}_{i} = \ind{i \leq w}$ and $w$ is the prescribed spectral window size. The filter retains the first $w$ components of the signal's DGFT, and we approximate the noisy  temperature signal by
\begin{equation} \label{e:filtered_signal}
\hat{\bbx} = \bbU \text{diag}(\tilde{\bbh})  \tilde{\bby} = \bbU \text{diag}(\tilde{\bbh}) \bbU^T \bby.
\end{equation}
Note that the filter $\bbH:=\bbU \text{diag}(\tilde{\bbh}) \bbU^T$ will in general not be expressible as a polynomial of the graph's adjacency matrix $\bbA$ (or some other graph shift operator~\cite{sandryhaila2013}), since DGFT modes need not be eigenvectors of the graph. Such a structure can be desirable to implement the filtering operation in a distributed fashion, and polynomial graph filter approximations of arbitrary lineal operators like $\bbH$ have been studied in~\cite{segarra2017filters}.

Fig.~\ref{fig:filtering_feasible}-(a) compares the original signal and the noisy measurements in the graph spectral domain induced by the DGFT. The original signal is low-pass bandlimited, compared to the noisy signal which spans a broader range of frequencies due to the white noise.  \blue{To better observe the low-pass property of the original signal, we also plot the cumulative energy of both the original and the noisy signals, defined by the percentage of the total energy present in the first $i$ frequency components for $i=1,\dots,N$. It is apparent from Fig.~\ref{fig:filtering_feasible}-(a) that the first few components of $\bbx$ capture most of its energy.}

Fig.~\ref{fig:filtering_feasible}-(b) shows a realization of the noisy graph signal $\bby$ superimposed with the denoised temperature profile $\hat{\bbx}$ obtained using \eqref{e:filtered_signal} with $w=3$, and the original signal $\bbx$. Filter design and the choice of $w$ is beyond the scope of this paper, but the
average recovery error $e_f = \| \hat{\bbx} - \bbx \|/\|\bbx\|$, over $1000$ Monte-Carlo simulations of independent noise,
\blue{attains a minimum of} approximately $12\%$ and Fig.~\ref{fig:filtering_feasible}-(b) shows $\hat{\bbx}$ closely approximates $\bbx$. 

 \blue{For this denoising task, we compare Algorithm~\ref{alg:feasible_dispersion} with a state-of-the-arte approach in~\cite{sardellitti}. The PAMAL algorithm~\cite[Algorithm~2]{sardellitti} fails to converge to an orthonormal basis within a reasonable time; however, we terminate the routine after $100$ iterations and work with the obtained basis. To assess the importance of the network model, we repeat the whole denoising experiment using a baseline contiguous US digraph (cf. Fig.~\ref{fig:temp_signal}), where we choose the direction of edges uniformly at random. To assess performance, we compute the relative recovery error with and without low-pass filtering as $e_f$ and $e=\lVert \bbn \rVert/\lVert \bbx \rVert$, respectively. Fig.~\ref{fig:filtering_feasible}-(c) depicts $e_f/e$ versus $w$ averaged over $1000$ Monte-Carlo simulations, which demonstrates the effectiveness of graph filtering in the dual domain enabled by the DGFT. Notice that our proposed approach outperforms the PAMAL method for both network models. Also, the minimum recovery error justifies the choice of $w=3$}. \blue{The results from the feasible method (S-N graph) are consistent with the energy plot, as the first $3$-$4$ components of $\bbx$ capture almost $99\%$ of the energy, and increasing the window size will only enhance the noise. As expected, Fig.~\ref{fig:filtering_feasible}-(c) shows that, on average, the performance degrades for the randomized US network using the fesible method, because the S-N digraph in Fig. \ref{fig:temp_signal} better captures the temperature flow.} While not shown here to avoid repetition, similar results with slightly higher recovery errors can be obtained using the greedy Algorithm~\ref{alg:greedy}. 

For additional comparisons we also \blue{consider} the following state-of-the-art \blue{GFT approaches for digraphs}: i)~filter design using adjacency matrix \cite{DSP_freq_analysis}; and ii)~the directed Laplacian in \cite{singh2016graph}}. We apply the filter design in \cite[Section V-B]{DSP_freq_analysis} to both the adjacency matrix and directed Laplacian defined as $\bbD_{\text{in}} - \bbA$, where $\bbD_{\text{in}}$ is the diagonal in-degree matrix. \blue{In the constructed S-N digraph, the adjacency matrix and directed Laplacian have only three and eight distinct eigenvalues, respectively. This limits the degrees of freedom for filter design, in which the low-pass graph filter is constructed through inverse polynomial interpolation using the distinct eigenvalues of the graph-shift operator. Such an ill-posed design gives rise to filters that amplify various high-frequency modes where only noise is present, resulting in large reconstruction errors. Furthermore, the Jordan decomposition required in \cite{singh2016graph} is  numerically unstable, an undesirable effect that  can be reduced using the Schur decomposition as proposed in~\cite{girault2015signal}. Although both the Jordan decomposition and the Schur-based block diagonalization preserve the subspaces for each eigenvalue, the degrees-of-freedom challenge remains as we only have a few distinct eigenvalues. Since the recovery errors obtained using~\cite[Section V]{DSP_freq_analysis} or~\cite{singh2016graph} are very variable and orders of magnitude higher than those reported in Fig.~\ref{fig:filtering_feasible}-(c) (e.g., the minimum average error over 1000 realizations is $4.58\times 10^4$ for the directed Laplacian), we have decided against plotting them to avoid hindering the clarity of the figure. Overall, these comparisons corroborate the merits of adopting spread (hence distinct) and orthogonal frequency modes, that also capture the notion of signal variation over a digraph.} 

\section{Conclusions}\label{S:Conclusions}
{We considered the problem of finding an orthonormal set of graph Fourier bases for digraphs. The starting point was to introduce a novel measure of directed variation to capture the notion of frequency on digraphs. Our DGFT design is to construct orthonormal frequency modes that take into account the underlying digraph structure, span the entire frequency range, and that are as evenly distributed as possible in the graph spectral domain to better capture notions of low, medium and high frequencies. To that end, we defined a spectral dispersion function to quantify the quality of any feasible solution compared to our ideal design, and minimized this criterion over the Stiefel manifold of orthonormal bases. To tackle the resulting non-convex problems, we developed two algorithms with complementary strengths to compute near-optimal solutions. First, we used a feasible method for optimization with orthogonality constraints, which offers provable convergence guarantees to stationary points of the spectral dispersion function. Second, we proposed a greedy heuristic to approximately minimize this dispersion using the eigenvectors of the Laplacian matrix of the underlying undirected graph. The greedy algorithm offers theoretical approximation guarantees by virtue of matroid theory and results for submodular function optimization. The overall DGFT construction pipeline is validated on a synthetic digraph with three communities as well as on a structural brain network. Finally, we show how the proposed DGFT facilitates the design of a low-pass filter used to denoise a real-world temperature signal supported on a network of the US contiguous states.

{With regards to future directions, }the complexity of finding {the} maximum frequency ($f_\text{max}$) on a digraph is an interesting open question. If NP-hard, it will be interesting to find the best achievable approximation factor (a $1/2$-approximation was given here). Furthermore, it would be valuable to quantify or bound the optimality gap for the stationary solution of the feasible method in the Stiefel manifold.

\appendix
\subsection{Feasibility of problem \eqref{eq:delta_opt_prob}}\label{S:s_feasibility}

The following proposition ensures that the spectral dispersion minimization problem \eqref{eq:delta_opt_prob} is feasible.

\begin{myproposition} \label{P:ortho_max_min}
The unit-norm basis vector $\bbu_{\max}$ defined in \eqref {e:opt_f_max} is orthogonal to the constant vector $\bbu_{\min}:=\frac{1}{\sqrt{N}}\mathbf{1}_N$.
\end{myproposition}
\begin{myproof}
	Since $\bbu_{\min}:=\frac{1}{\sqrt{N}}\mathbf{1}_N$, we will show that $\bbu_{\max}^T\mathbf{1}_N=0$. Arguing by contradiction, suppose that the sum of the entries in $\bbu_{\max}$ is not zero. We show that $\text{DV}(\bbu_{\max})$ can be improved in that case, which contradicts the optimality of $\bbu_{\max}$. 
	
	Without loss of generality assume that $\bbu_{\max}^T\mathbf{1}_N=\epsilon>0$, and define $\bar{\bbu}:= \bbu_{\max}-\frac{\epsilon}{N}\mathbf{1}_N$. First, note that $\text{DV}(\bbu_{\max})=\text{DV}(\bar{\bbu})$, since adding (subtracting) a constant to (from) all coordinates will not change the directed variation. Second, 
	%
	\begin{equation*}
		\|\bar{\bbu}\|^2 
		= \bbu_{\max}^T \bbu_{\max}-\frac{2\epsilon}{N} \bbu_{\max}^T\mathbf{1}_N+\left(\frac{\epsilon}{N}\right)^2\mathbf{1}_N^T\mathbf{1}_N 
		=  1-\frac{\epsilon^2}{N}. 
	\end{equation*}
	Therefore, we have a new vector $\bar{\bbu}$ with the same directed variation but smaller norm. Now we can scale this vector as $\alpha \bar{\bbu}$ (with $\alpha>1$) to obtain a normalized vector with $\text{DV}(\alpha \bar{\bbu})=\alpha^2 \text{DV}(\bar{\bbu})$, which improves upon $\bbu_{\max}$.
\end{myproof}

\subsection{Proof of Proposition \ref{pro:special_cases}-1)} \label{S:s_path_proof}
We prove by induction (on the length of path) that the maximum frequency on a dipath is twice the maximum edge weight. Let \blue{$x_1,x_2,...,x_N$ be the signal values} on the dipath of length $N-1$, with directed \blue{edges going from $i$ to $i+1$, $i=1,...,N-1$.} For the base case of $N=2$, we have to maximize $A_{12}(x_1-x_2)^2$ subject to $x_1\geq x_2$ and $x_1^2+x_2^2=1$. The solution to this optimization problem is $x_1=\sqrt{2}/2$ and $x_2=-\sqrt{2}/2$ which evaluates to $A_{12}(x_1-x_2)^2=2A_{12}$. For the inductive step, assume that the claim is true for a dipath of length $N-1$. We show that it should be the case for $N$ edges as well. If $x_N\leq x_{N+1}$ in the optimal solution for $N$ edges, then for the last edge $[x_N-x_{N+1}]_{+}=0$ and the optimal directed variation is obtained from the first $N-1$ edges, which is twice their largest edge weight by assumption. Indeed, note that $A_{N(N+1)}$ cannot be the maximum edge weight in this case, otherwise setting $x_N=\sqrt{2}/2$ and $x_{N+1}=-\sqrt{2}/2$ would improve the optimal solution and violates the assumption of $x_N\leq x_{N+1}$.  \blue{Next}, we assume $x_N>x_{N+1}$. In this case, we claim that $x_N$ should also be greater than or equal to $x_{N-1}$. If not, we have $x_{N-1}> x_N>x_{N+1}$, which cannot be an optimal solution. To see this, we can swap the value of $x_N$ with either $x_{N-1}$ or $x_{N+1}$ and improve the directed variation because either $A_{(N-1)N}(x_{N-1}-x_{N+1})^2$ or $A_{N(N+1)}(x_{N-1}-x_{N+1})^2$ is greater than $A_{(N-1)N}(x_{N-1}-x_N)^2+A_{N(N+1)}(x_{N}-x_{N+1})^2$. Finally, if $x_N>x_{N+1}$ and $x_{N}\geq x_{N-1}$, then the edge $({N-1},N)$ does not contribute to the directed variation and the path is divided into two sections. Since both $\| \bbu \|^2$ and $\text{DV}(\bbu)$ scale quadratically with $\bbu$, one can show (in general) that one of the optimal solutions should be achieved by only the variations of a set of connected edges; otherwise it is better to void the section with the lower ratio of objective to norm, and scale up the other section. This means that in our dipath example, once the edge $({N-1},N)$ has zero objective, we can also make one of the two sections zero. The claim is then true by the inductive assumption. The achievability of the maximum directed variation follows by setting $\pm \sqrt{2}/2$ on the edge with largest weight.\hfill$\blacksquare$

\subsection{Proof of Proposition \ref{pro:special_cases}-2)} \label{S:s_cycle_proof}
There should be at least one edge $(i,j)$ in the cycle for which $[x_i-x_j]_{+}=0$, otherwise we obtain a closed loop of strict inequalities among consecutive $x_i$ values which is impossible. Given that edge $(i,j)$ has zero directed variation, the rest of the cycle can be viewed as a dipath which has directed variation of at most $2$ times the largest edge weight by Proposition~\ref{pro:special_cases}-1). The same argument ensures the achievability of the solution.\hfill$\blacksquare$

\subsection{Proof of Proposition \ref{pro:special_cases}-3)} \label{S:s_bipartite_proof}
Let $\ccalV^{+}$ and $\ccalV^{-}$ be the two node partitions of the unidirectional bipartite graph, where the edges are constrained to go from $\ccalV^{+}$ to $\ccalV^{-}$. First, we show that in the optimal solution maximizing the directed variation we must have $x_i\geq 0$ for all $i \in \ccalV^{+}$, and $x_i\leq 0$ for all $i \in \ccalV^{-}$. Otherwise, assume that there exists some node $j\in \ccalV^{-}$ with $x_j>0$. Then we can improve the directed variation by setting $x_j=0$, because $j$ has only incoming edges and decreasing $x_j$ will not decrease the variation on such edges (and we gain some slack in the norm constraint by this change). Similarly, we arrive at a contradiction if some node in $\ccalV^{+}$ has negative value.

With this information, we know that all the summands $A_{ij}[x_i-x_j]_{+}^2$ in the objective are indeed equal to $A_{ij}(x_i-x_j)^2$, because $i\in \ccalV^{+}$, $j\in \ccalV^{-}$, and $x_i\geq 0\geq x_j$. Therefore, we can replace the cost function with the total variation and solve the following optimization problem instead
\begin{equation}
\label{eq:bipartite_f_max}
\begin{aligned}
& \max_{\bbx}
& & \text{TV}(\bbx)=\bbx^T \bbL \bbx \\
& \text{subject to}
& & \bbx^T\bbx=1\\
&
&& x_i\geq 0, \quad \quad i\in \ccalV^{+}\\
&
&& x_j\leq 0, \quad \quad j\in \ccalV^{-}.
\end{aligned}
\end{equation}
Assume that we relax problem (\ref{eq:bipartite_f_max}) by dropping the inequality constraints. Once we do that, the solution will be $\lambda_{\max}$. The next lemma shows this relaxation entails no loss of optimality.
\begin{mylemma}\label{lemma_appendix}
\normalfont For an undirected bipartite graph $\ccalG=(\ccalV_1,\ccalV_2,\bbA)$ with Laplacian matrix $\bbL$, let $\bbu\in\reals^N$ be the dominant eigenvector of $\bbL$ (corresponding to $\lambda_{\max}$). Then $\bbu$ has the same sign over the coordinates of each partition (i.e., non-negative for $\ccalV_1$ and non-positive for $\ccalV_2$ or vice versa).
\end{mylemma}
\begin{myproof}
\normalfont
If by contradiction that were not the case, we could change the signs (maintaining the absolute values and vector norm) to be positive in one partition (say $\ccalV_1$) and negative in the other ($\ccalV_2$). This change does not decrease the absolute difference between signal values on nodes incident to each edge, and contradicts the fact that $\bbu$ maximizes $\text{TV}(\bbu)=\bbu^T \bbL \bbu=\sum_{i,j=1,j>1}^N{A_{ij}(u_i-u_j)^2}$.
\end{myproof}

By virtue of Lemma \ref{lemma_appendix}, either $\bbu$ or $-\bbu$ is feasible in \eqref{eq:bipartite_f_max} and attains the optimum objective value $\lambda_{\max}$.\hfill$\blacksquare$

\bibliographystyle{IEEEtranS}
%
\bibliography{refs}

\end{document}